\documentclass[manuscript,screen]{acmart}

\usepackage{ising_v1}
\usepackage{algorithmic}
\usepackage{graphicx}
\usepackage{textcomp}
\usepackage{xcolor}
\usepackage{tabularx}
\usepackage{multirow}

\def\submission{1}

\renewcommand\footnotetextcopyrightpermission[1]{}

\acmJournal{CSUR}



\begin{document}

\title{SoK: Fully Homomorphic Encryption Accelerators}


\author{Junxue Zhang}
\authornote{Both authors contributed equally to this research.}
\email{zjx@cse.ust.hk}
\orcid{0000-0001-6926-7801}
\affiliation{%
  \institution{iSINGLab @ HKUST}
  \country{Hong Kong, China}
}
\affiliation{%
  \institution{Clustar}
  \country{China}
}

\author{Xiaodian Cheng}
\authornotemark[1]
\email{xchengaq@connect.ust.hk}
\affiliation{%
  \institution{iSINGLab @ HKUST}
  \country{Hong Kong, China}
}

\author{Liu Yang}
\email{lyangau@connect.ust.hk}
\affiliation{%
  \institution{iSINGLab @ HKUST}
  \country{Hong Kong, China}
}
\affiliation{%
  \institution{Clustar}
  \country{China}
}

\author{Jinbin Hu}
\email{jinbinhu@ust.hk}
\affiliation{%
  \institution{iSINGLab @ HKUST}
  \country{Hong Kong, China}
}

\author{Ximeng Liu}
\email{snbnix@gmail.com}
\affiliation{%
  \institution{Fuzhou University}
  \country{China}
}

\author{Kai Chen}
\email{kaichen@cse.ust.hk}
\affiliation{%
  \institution{iSINGLab @ HKUST}
  \country{Hong Kong, China}
}

\begin{abstract}

Fully Homomorphic Encryption~(FHE) is a key technology enabling privacy-preserving computing. However, the fundamental challenge of FHE is its inefficiency, due primarily to the underlying polynomial computations with high computation complexity and extremely time-consuming ciphertext maintenance operations. To tackle this challenge, various FHE accelerators have recently been proposed by both research and industrial communities. This paper takes the first initiative to conduct a systematic study on the 14 FHE accelerators --- cuHE/cuFHE, nuFHE, HEAT, HEAX, HEXL, HEXL-FPGA, 100$\times$, F1, CraterLake, BTS, ARK, Poseidon, FAB and TensorFHE. We first make our observations on the evolution trajectory of these existing FHE accelerators to establish a qualitative connection between them. Then, we perform testbed evaluations of representative open-source FHE accelerators to provide a quantitative comparison on them. Finally, with the insights learned from both qualitative and quantitative studies, we discuss potential directions to inform the future design and implementation for FHE accelerators.
\end{abstract} 

\maketitle

\thispagestyle{plain}
\pagestyle{plain}

\section{Introduction}
\label{sec:introduction}

With the increasing concern about data privacy and integrity, privacy-preserving computing has been adopted in many real-world applications, \eg, cloud computing~\cite{sphinx}, machine learning~\cite{federated_learning,secureml}, database search~\cite{onion_pir}, \etc. Among all the privacy-preserving technologies, fully homomorphic encryption~(FHE) emerges as one of the most important and promising technologies and has been adopted in various applications~\cite{poseidon,onion_pir,resnet20,fhe_psi}. Specifically, FHE allows performing arbitrary operations directly over the encrypted data without decryption, making it appealing for privacy-preserving computation.


Although promising, one fundamental drawback of FHE is its \emph{inefficiency}. Compared to plaintext computation, FHE-enabled computation is orders of magnitude slower, which restricts its deployment in many performance-critical systems. To solve this problem, various optimizations have been proposed. 
One direction is to improve the efficiency of the algorithm. For example, modern FHE schemes, such as BGV~\cite{bgv}, BFV~\cite{bfv}, and CKKS~\cite{ckks}, all support SIMD-like~(\ie, batching) operations~\cite{simd} to pack many plaintext messages into one ciphertext to improve the execution efficiency.  
Instead of studying these algorithm-level performance optimizations, in this paper we focus on the other key direction --- leveraging hardware accelerators to improve the efficiency of FHE schemes\footnote{Note that algorithm optimization and hardware acceleration are complementary, and can be combined to improve the overall performance.}.

Before introducing FHE accelerators, we first illustrate what makes FHE slow and the challenges of acceleration. In this paper, we find that the root cause of FHE's inefficiency is two-fold: underlying polynomial computations with high computation complexity and two extremely time-consuming ciphertext maintenance operations. First, most of FHE's underlying operations are polynomial operations, which are much more complex than plaintext computation, where operands are integers or floating numbers. While Fast Fourier Transform~(FFT)/Number Theoretic Transform~(NTT) can be utilized to speed up the polynomial operations~\cite{intro_to_alg} from algorithm-level, further accelerating NTT/FFT faces challenges in three aspects: high computation complexity, extremely intensive memory access, and limited generality~(\S\ref{sec:challenge_polynomial}). Second, compared to plaintext computation, FHE requires ciphertext maintenance operations to ensure correctness. Such operations involve over-complicated computation steps, causing further performance degradation. Although they are built upon polynomial operations, fully accelerating ciphertext maintenance operations is more challenging in terms of the aforementioned three aspects compared to merely accelerating polynomial operations~(\S\ref{sec:challenge_keyswitching} and \S\ref{sec:challenge_bootstraping}).


To improve the efficiency of FHE, FHE accelerators are proposed. Initially, these accelerators rely on features provided by general hardware. For example, Intel proposed Intel Homomorphic Encryption Acceleration Library~(HEXL) to leverage AVX-512 instructions for fast NTT operations~\cite{hexl}. nuFHE~\cite{nufhe-url} and 100$\times$~\cite{over100} used GPU implementations, \ie, CUDA~\cite{cuda-url} programs, to accelerate TFHE~\cite{tfhe} and CKKS~\cite{ckks} respectively. While providing notable acceleration for FHE schemes, these accelerators are far from satisfactory.

To further improve the performance of FHE schemes, people begin to exploit specific-designed hardware accelerators. Field Programmable Gate Array~(FPGA) is first used to build circuits to efficiently execute NTT, inverse NTT~(iNTT), and key-switching in FHE schemes~\cite{hexl-fpga-url,heax,fpga_fhe_hpec,fpga_fhe_tc,fpga_fhe_tecs,fpga_fhe_tetc,fpga_fhe_vlsi}. These FPGA-based accelerators are affordable but suffer from intrinsic disadvantages of FPGA itself: limited programmable resources and low working frequency~\cite{gap_fpga_asic}. Then, to overcome these disadvantages, people are seeking expensive Application-specific Integrated Circuit~(ASIC) technologies to build high-performant FHE accelerators~\cite{f1,craterlake,bts,ark}. While the acceleration ratios of these ASIC-based accelerators are promising, \eg, $\sim 1000\times$ and $\sim 100\times$ NTT throughput compared to FPGA-based and GPU-based solutions respectively, they are way more expensive. For instance, developing and taping out a 12nm ASIC like~\cite{craterlake} requires millions of US dollars.

In the recent decade, we have seen an explosive growth of FHE accelerators~\cite{heaws,heax,hexl-fpga-url,f1,craterlake,bts,fpga_fhe_hpec,fpga_fhe_tc,fpga_fhe_tecs,fpga_fhe_tetc,fpga_fhe_vlsi,ark}, and expect an increasingly more active development of FHE accelerators in the near future. However, we lack a comprehensive and systematic study to shed light on the status quo of existing FHE accelerators, which could inspire the future design and implementation of FHE accelerators. 

Motivated by this, we take the first initiative to perform the systematization of knowledge on FHE accelerators. We first review 14 existing FHE accelerators and make observations on the evolution trajectory of these works, which establishes a qualitative connection among them~(\S\ref{sec:fhe_acc_sok}). Then, to give readers a clear view of how these accelerators perform, we present a quantitative analysis of them. Specifically, we use testbed experiments to evaluate the performance of some representative open-source accelerators, and further include the statistics from papers of other well-known but closed-source accelerators \bluehighlight{for thoroughness}~(\S\ref{sec:evaluation}). Finally, based on our qualitative and quantitative analysis, we discuss the potential future directions, such as new design tradeoffs, software/hardware co-designs, scaling methodologies, \etc, to inform the future design and implementation of FHE accelerators~(\S\ref{sec:discussion}).



Along with the paper, we provide a Docker image\footnote{\url{https://hub.docker.com/r/hpfilter/sok-fhe-accelerator}} that includes all configurations and scripts for performance evaluations of all open-sourced FHE accelerators used in our paper, which can be readily reused by the community.

\parab{Related Works:} Systematization of knowledge on FHE library~\cite{survey_he} and compiler~\cite{sok_fhe_compiler} have already been proposed. To the best of our knowledge, our paper is the first systematization of knowledge on FHE accelerators. FHE accelerators are closely related to FHE libraries and compilers. For example, some FHE accelerators are designed to work with particular FHE libraries, \eg, HEAX chooses SEAL~\cite{seal-url} as its target library to accelerate. Moreover, an increasing number of FHE accelerators use compilers for a software/hardware co-design~\cite{f1,bts,craterlake,ark}. However, few FHE accelerators consider leveraging existing FHE compilers, such as EVA~\cite{eva}, E3~\cite{e3}, \etc, which, in our opinion, still leaves dramatic design space for better performance and flexibility.
\bluehighlight{
\section{Preliminaries}
\label{sec:preliminary}

In Table \ref{tab:notations}, we list the notations used in the paper. Here, we define a polynomial $A$ as follows:

\begin{equation}
    A(x) = \sum_{j=0}^{n-1} a_j x^j
\end{equation}

The degree of a polynomial is the highest power of the variable $x$ with a non-zero coefficient. Integer $n$ is defined as the degree-bound of the polynomial, which is strictly larger than the degree of the polynomial~\cite{intro_to_alg}. In many previous works, $n$ is also called the degree of a polynomial for simplicity. In this paper, we use the term degree rather than degree-bound to refer to $n$.

\subsection{Ciphertexts and Keys}
\label{sec:cipherandkey}
In this paper, all the cryptosystems are asymmetric, which means we have encrypt plaintext $m$ with public key $\boldsymbol{\bf pk}$ and decrypt ciphertext $\boldsymbol{\bf ct}$ with secret key $s$. All the ciphertexts and keys in the algorithms covered in the paper are in the form of polynomials. Therefore, the underlying operations of FHE schemes are all polynomial operations.

\subsection{Polynomial Operations} 
\label{sec:FFTandNTT}
As discussed, polynomial operations, including polynomial additions and multiplications, are the basic building blocks of the FHE algorithms.
To implement the operations, there are two common representations of a polynomial: coefficient representation and point-value representation. The polynomials in FHE are naturally stored in the coefficient representation. However, the time complexity of multiplication between polynomials in the coefficient representation is $O(n^2)$, while it can be reduced to $O(n)$ with the point-value representation. A popular approach for representation conversion is the Fast Fourier Transform, which leverages the idea of divide-and-conquer to reduce the conversion time complexity to $O(n\log n)$. Note that FFT only works on complex numbers. For RLWE-based FHE schemes, we also need Number Theoretic Transform, which is a generalization of FFT but works over finite fields. Readers may refer to Appendix~\ref{app:ntt} for more details.

\subsection{RNS Decomposition} 
\label{sec:rns}
FHE schemes are mainly constructed over polynomial rings, which means we perform polynomial operations with the coefficients modulus $Q$ and $P$. $Q$ and $P$ are large integers chosen for ciphertext calculation and key generation, respectively. However, polynomial operations with large integers lead to performance degradation~\cite{ckks}. For performance optimization, a common strategy is to use Residue Number System~(RNS). RNS decomposes modulus $Q$ and $P$ into the product of several smaller coprime moduli $q_0\cdot q_1\cdot \dots \cdot q_L$ and $p_0\cdot p_1\cdot \dots \cdot p_\alpha$, allowed by the Chinese Remainder Theorem ~(CRT)~\cite{crt}. Multiple polynomials with smaller moduli are used to replace the original polynomial. By applying RNS to FHE, the coefficient size of polynomials is greatly reduced at the cost of decomposing each polynomial into multiple ones. Since the calculation complexity of modular multiplication is roughly proportional to the square of the coefficients' bit width, the overall complexity is reduced. In word-wise FHE algorithm, the value $L$ is also called the multiplicative depth of a ciphertext as the moduli $q_0\cdot q_1\cdot \dots \cdot q_L$ are gradually consumed by the homomorphic multiplications and $L$ usually determines the maximum number of multiplications that can be supported by the FHE ciphertexts.

}
\section{Fully Homomorphic Encryption}
\label{sec:fhe_background}

\begin{table}[t]
\small
\renewcommand\arraystretch{1.2}
\centering
\begin{tabularx}{0.6\linewidth}{l l}
\toprule
\bf Notation & \bf Definition \\
\midrule
$n$ & Degree of a polynomial. \\
$m$ & Plaintext. \\
$\boldsymbol{\bf ct}$ & Ciphertext. \\
$s$ & Secret key. \\
$\boldsymbol{\bf pk}$ & Public key. \\
$Q$ & Coefficient modulus of ciphertext polynomial. \\
$P$ & Special modulus for the keys. \\
$\{q_i, i\in[0,L]\}$ & A set of moduli. $Q=\prod_{i=0}^Lq_i$. \\
$\{p_i,i\in[0,\alpha-1]\}$ & A set of special moduli. $P=\prod_{i=0}^{\alpha-1} p_i$. \\
$L$ & Multiplicative depth of a fresh ciphertext.  \\
$l$ & Current multiplicative depth of a ciphertext.  \\
${\rm dnum}$ & Decomposition number in key-switching. \\
$\alpha$ & $\#$ of special moduli $p_i$. $\alpha=\lfloor (L+1)/{\rm dnum}\rfloor $. \\
$\{Q_j,j\in[0,{\rm dnum}]\} $ & A set of modulus factors. $Q_j=\prod_{i = j\alpha}^{(j+1)\alpha-1}q_i$. \\
$\boldsymbol{\bf swk}$ & Switching key for key-switching. \\
$q$ & Coefficient modulus of plaintext polynomial. \\
$\Delta$ & Scaling factor. \\
\bottomrule
\end{tabularx}
\caption{Notations used in this paper.}
\label{tab:notations}
\end{table}

Homomorphic encryption is an encryption scheme that allows performing arbitrary computation over ciphertexts without decrypting them. For example, Paillier is an additive homomorphic encryption scheme thus we can perform additions over ciphertext~\cite{paillier}, while RSA~\cite{rsa} is a multiplicative homomorphic encryption scheme. In this paper, we are focusing on fully homomorphic encryption~(FHE) schemes\footnote{Similar to \cite{sok_fhe_compiler}, FHE schemes in our paper include leveled FHE.}, which allow both additive and multiplicative homomorphic operations.

Current FHE schemes can be categorized into word-wise FHE and bit-wise FHE~\cite{pegasus}. Word-wise FHE supports algebraic operations on word-based~(or message-based) encrypted data. Moreover, word-wise FHE supports efficient single-instruction-multiple-data~(SIMD) style homomorphic operations, \eg, performing homomorphic addition and multiplication over batched plaintext~\cite{simd}. However, word-wise FHE is not suitable for evaluating non-polynomial functions, \eg, sigmoid/relu functions, which are commonly used in machine learning applications. Examples of word-wise FHE are CKKS~\cite{ckks}, BFV~\cite{bfv}, BGV~\cite{bgv}, \etc. In contrast, bit-wise FHE supports operations of boolean circuits. As its name indicates, bit-wise FHE schemes encrypt each bit of the plaintext and are usually used to evaluate non-polynomial operations by constructing lookup tables. Bit-wise FHE schemes do not provide sufficient support for SIMD-style operations and usually suffer from a higher degree of ciphertext inflation, which poses a larger challenge to memory bandwidth. Examples of bit-wise FHE schemes are TFHE~\cite{tfhe} and FHEW~\cite{fhew}. In order to support real-world privacy-preserving applications, such as privacy-preserving machine learning, both word-wise and bit-wise FHE schemes are used together~\cite{pegasus,chimera}.

\begin{figure}[t!]
\centering
\includegraphics[width=0.9\columnwidth]{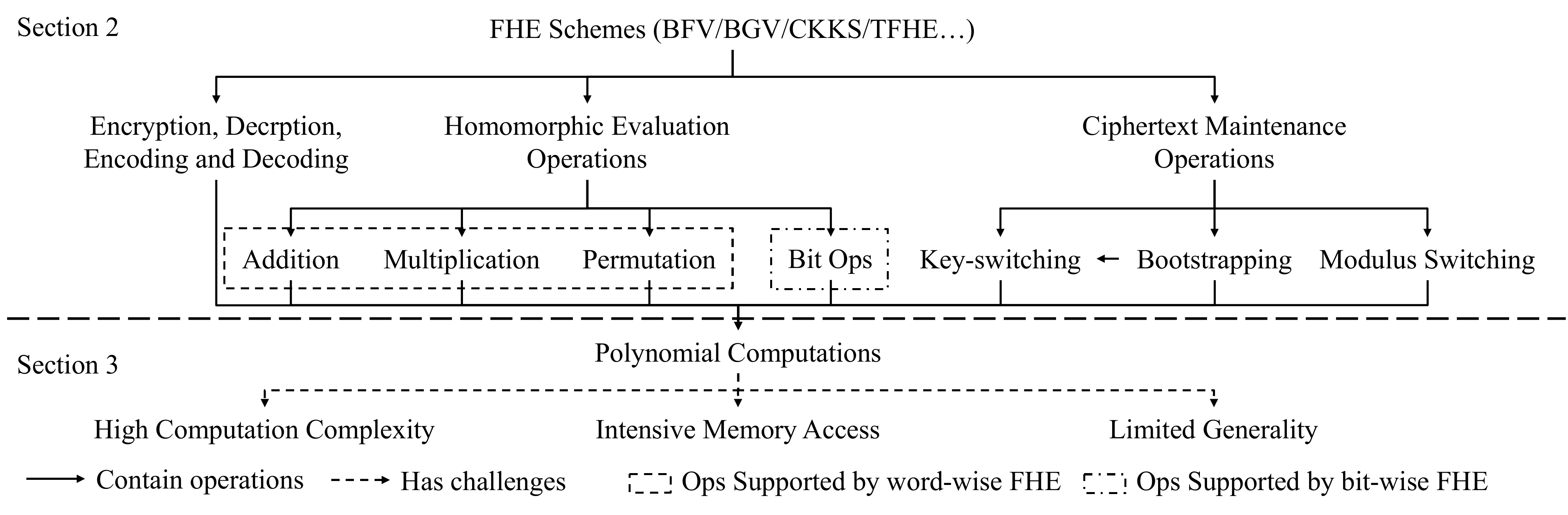}
\caption{\bluehighlight{Overview of the operations used in FHE schemes.}}
\label{fig:background-overview}
\end{figure}

In this article, we focus on word-wise FHE algorithms constructed over ring learning with errors problem~(RLWE)~(\eg, BFV, BGV and CKKS)~\cite{rlwe} and bit-wise FHE algorithm (\eg, TFHE)~\cite{gsw}
because they are practical and widely adopted. Since these FHE schemes mainly manipulate polynomials over finite fields or torus, most of the operations to be discussed later are made up of polynomial computations. In the following sections, we will show them in detail. Figure~\ref{fig:background-overview} demonstrates the relationship of operations of FHE.

\subsection{Encoding, Decoding, Encryption \& Decryption}
\label{sec:encoding}
\parab{Encoding and Decoding:} The goal of encoding is to convert plaintext messages into polynomials for the subsequent homomorphic operations. 
Please note that by packing a vector of numbers during encoding, specific FHE schemes, such as BGV and BFV, can naturally support SIMD-style operations. Conversely, decoding is used to recover plaintext messages from polynomials.

\parab{Encryption:} In an asymmetric encryption system, a ciphertext $\boldsymbol{\rm ct}$ is generated by obfuscating the plaintext with a public key and noises. For example, in RLWE-based FHE, the public key is an RLWE instance, $\boldsymbol{\rm pk}=({\rm pk_0}, {\rm pk_1})=(as+e, a)$, generated from the secret key $s$. The coefficients of random polynomial $a$ are uniformly sampled from interval $(-Q/2, Q/2]$, and coefficients of noise polynomial $e$ are independently sampled from a discrete Gaussian distribution. The encryption is formulated as 
\begin{equation}
\boldsymbol{\rm ct}=(c_0,c_1)=v\cdot \boldsymbol{\rm pk} + (\Delta \cdot m + e_1, e_0) \mod Q,
\end{equation}
where $m$ is the plaintext (\ie, encoding result), $e_0$ and $e_1$ are also noise polynomials, $v$ is a random polynomial with small coefficients, and $\Delta$ is the scaling factor in some schemes to control precision.

\parab{Decryption:} Although different FHE schemes use different decryption workflows, the common objective is to recover the plaintext by removing $a \cdot s$ from the ciphertext. As a result, they share the same core operation in decryption, which is $c_0-c_1\cdot s$.

\subsection{Homomorphic Evaluation}
\label{sec:homomorphic_evaluation}
For word-wise FHE, major homomorphic evaluations include multiplications, additions, subtractions and permutations, which are sufficient for most applications. Non-polynomial functions can also be achieved via polynomial approximation, which covers a large range of applications. Bit-wise FHE schemes can efficiently evaluate bit operations, including NOT, AND, NAND, OR and XOR, which can provide accurate results of non-polynomial functions.

However, homomorphic evaluations lead to two significant problems. First, operations such as multiplication and permutation construct special ciphertexts that cannot be directly decrypted or used as the input of subsequent operations. Second, as noise is introduced to secure the ciphertext in FHE schemes, it gradually grows during FHE operations, especially in homomorphic multiplications. After the noise exceeds a threshold, it will impact the correctness of the decryption. Therefore, ciphertext maintenance operations, \eg, key-switching, modulus-switching and bootstrapping, are required in FHE to solve these problems.

\subsection{Ciphertext Maintenance}
\label{sec:ct_maintenance}
\parab{Key-switching:} As its name implies, key-switching homomorphically switches the secret key of a ciphertext while keeping the corresponding plaintext unchanged. More specifically, ${\boldsymbol{\rm ct}}$ is the ciphertext of plaintext $m$, and it can be decrypted with some special secret key $s'$. After executing ${\boldsymbol{\rm ct'}}={\rm Keyswitch}({\boldsymbol{\rm ct}}, \boldsymbol{\rm swk})$, ciphertext ${\boldsymbol{\rm ct'}}$ can be decrypted with the original secret key $s$ and the corresponding plaintext is still $m$. In the equation, $\boldsymbol{\rm swk}$ is a pre-generated public key called switching key, and it can be considered a ciphertext of $P\cdot s'$ with modulus $P\cdot Q$. $P$ is an integer used to control the scale of noise in key-switching. 
After certain operations such as homomorphic multiplication and permutation, key-switching is leveraged to convert the resulting ciphertexts back into the original form for the following operations. Therefore, it is intensively used in FHE schemes.

\parab{Modulus Switching and Multiplicative Depth:} Generally speaking, modulus switching refers to the operations that switch the modulus of a ciphertext. It is largely applied in FHE schemes to raise or reduce the modulus for different purposes. In particular, RLWE-based FHE schemes introduce modulus switching to control the proportion of noise in the ciphertext at the cost of reducing the modulus $Q$. When $Q$ is too small to support further operations, the noise cannot be reduced anymore. Consequently, the size of modulus $Q$ limits the number of consecutive homomorphic operations on a freshly encrypted ciphertext. The number is also called the maximum multiplicative depth~(or budget) $L$ of arithmetics supported by the FHE scheme. Such FHE schemes are usually called leveled fully homomorphic encryption~(leveled FHE) because of the limitations on $L$. In this paper, we use the term deep and shallow to describe applications that consume large multiplicative depth ~(\eg, deep neural networks~\cite{resnet20}) and those that only contain a few multiplications ~(\eg, database lookup~\cite{dblookup}), respectively. Leveled FHE schemes are designed to be efficient for shallow computations. In deep applications, excessive $Q$ dramatically reduces the performance. Bootstrapping is leveraged to refresh the ciphertext and recover the multiplicative depth, which we will discuss next.

\parab{Bootstrapping:}
Bootstrapping is a generic term for operations that refresh a ciphertext in FHE. Their common idea is to homomorphically recrypt the old ciphertext and generate a fresh one.

In bit-wise FHE schemes such as TFHE, bootstrapping is performed in every bit-wise operation. Therefore, multiplicative depth is not considered in TFHE. The most time-consuming part of bootstrapping in TFHE is the homomorphic evaluation of a lookup table~(LUT). The lookup process can be implemented through a large number of multiplexer~(MUX) gates. Since the MUX gate mainly consists of polynomial additions, subtractions and multiplications, polynomial computations are the major workload in TFHE.

Unlike bit-wise FHE, bootstrapping in RLWE-based word-wise FHE schemes is more complicated. We tend to reduce the frequency of bootstrapping and perform one only when the current multiplicative depth is not enough. It is worth noting that bootstrapping combines multiple operations, including many multiplications, thus consuming considerable multiplicative depth by itself. The concrete implementations of bootstrapping are not the same in various word-wise FHE algorithms, but their workflow can be summarized into four common steps.

\noindent Step 1. Modulus Switching: In word-wise FHE, the multiplicative depth is proportional to the modulus size of the ciphertext. Therefore, the modulus should be raised if more multiplications are needed. Given ciphertext ${\rm ct}$ under modulus $Q$, the first step in bootstrapping is to generate a new ciphertext ${\rm ct'}$ encrypting the same plaintext while extending the modulus to $Q'$ which satisfies $Q'\gg Q$.

\noindent Step 2. CoeffToSlot: Although the modulus of the ciphertext has been raised, the decryption formula no longer holds since the coefficients of the plaintext polynomial are not guaranteed to be bounded by the modulus of the plaintext during the process of modulus switching. In this case, homomorphic evaluations are required to modulo the coefficients. However, we can only operate on plaintext slots rather than polynomial coefficients with homomorphic operations in FHE. To make the coefficients accessible for homomorphic evaluations, homomorphic encoding should be performed in advance to put the coefficients of plaintext polynomial into the plaintext slots. This process is also called CoeffToSlot~\cite{ckks-bootstrapping, better-bootstrapping} or linear transformation~\cite{helib-bootstrapping, improved-fhe-boostrapping}. 

\noindent Step 3. Homomorphic Evaluation: Homomorphic evaluations in bootstrapping are introduced to modulo the coefficients of plaintext polynomial. They contain non-polynomial functions that can not be directly executed in FHE. For example, in BGV and BFV, the main operation of this step is digit extraction, while modulus reduction is the dominant operation of this step in CKKS. Thus, in modern FHE implementations, approximation schemes such as Taylor expansion~\cite{ckks-bootstrapping} and optimized Chebyshev method~\cite{better-bootstrapping,improved-fhe-boostrapping} are used to alternatively evaluate high-degree polynomials.

\noindent Step 4. SlotToCoeff: After homomorphic evaluation, the results in the plaintext slots should be placed back to the coefficients of the plaintext polynomial. The process is an inverse operation of CoeffToSlot, which is called SlotToCoeff or inverse linear transformation.

\section{What \bluehighlight{Makes} FHE Slow \& the Challenges of Accelerating Them}
\label{sec:fhe_inefficiency}

FHE is an ideal solution for many privacy-preserving applications since it can simultaneously protect confidential data and satisfy emerging data protection lawsuits and regulations~\cite{gdpr-url}. However, FHE still suffers from inefficiency, which is the focus of the paper. It is worth noting that previous works have also mentioned several other reasons restricting the broad adoption of FHE, such as its usage complexity~\cite{sok_fhe_compiler}. However, we believe the inefficiency problem is still the major roadblock to FHE's adoption in the production environment.


In this paper, we have identified the cause of FHE's inefficiency as a two-fold problem. First, most of FHE's operations are built upon polynomials~\cite{bfv,bgv,ckks,tfhe}. Therefore, FHE is inefficient since polynomial computations, specifically polynomial multiplication, are naturally much more complex than integer/floating number calculations. Second, two necessary ciphertext maintenance operations, \ie, key-switching and bootstrapping, are extremely time-consuming since they involve very complicated computations. As reported in ARK~\cite{ark}, the two major components~(\ie, NTT and fast basic conversion, which will be introduced in detail later) in key-switching take up more than $80\%$ of the total computation time. According to the analysis in~\cite{craterlake}, bootstrapping may take up over $90\%$ of computation time in an end-to-end FHE task~(bootstrapping consists of key-switching operations).


To solve the problems, FHE accelerators are proposed. They target improving the performance of FHE schemes by either leveraging general hardware~(SIMD feature provided by CPUs) or relying on completely specific hardware~(FPGA-based circuits). However, it is not easy for these accelerators to achieve ideal performance. In the following part, we will first summarize the challenges of accelerating polynomial computation~(specifically, its core operation: NTT/FFT) into three aspects: \emph{high computation complexity, extremely intensive memory access and limited generality}. Second, we will further demonstrate the challenges of accelerating key-switching and bootstrapping. Although these two operations are built upon polynomials operations, designing an end-to-end acceleration solution for them is more challenging in terms of the aforementioned three aspects, which makes it increasingly more difficult to design efficient FHE accelerators.

\subsection{Challenges of Accelerating Polynomial Computation}
\label{sec:challenge_polynomial}
One widely-adopted algorithm-level optimization for polynomial computation is to leverage Fast Fourier Transform~(FFT)/Number Theoretic Transform~(NTT)~\cite{seal-url,palisade-url}. Thus, accelerating FFT/NTT is essential to all FHE accelerators. In the following part, we will present the three challenges of further accelerating FFT/NTT with the current hardware architecture.

\subsubsection{High Computation Complexity}
\label{sec:ntt_complexity}
\begin{figure*}[t]
\begin{minipage}[t]{.40\textwidth}\vspace{0pt}\raggedleft
\begin{minipage}[t]{\columnwidth}\vspace{0pt}\raggedleft
\includegraphics[width=\columnwidth]{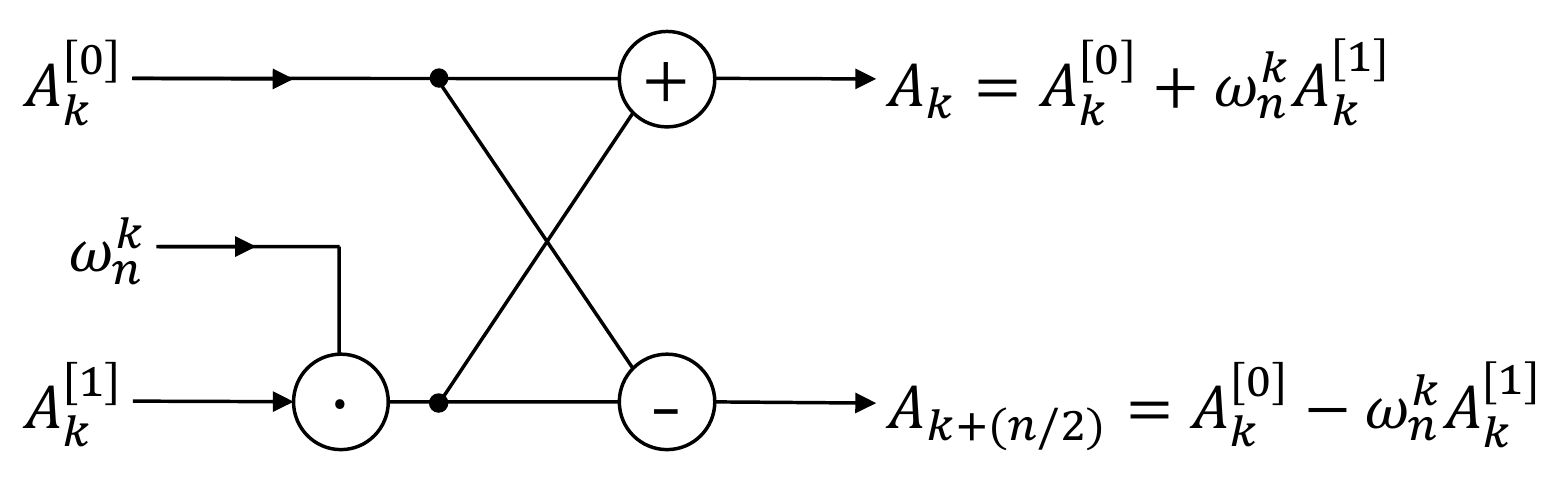}
\caption{Cooley-Tukey butterfly}
\label{fig:ct-butterfly}
\end{minipage}
\begin{minipage}[t]{\textwidth}\vspace{15pt}\raggedleft
\includegraphics[width=\columnwidth]{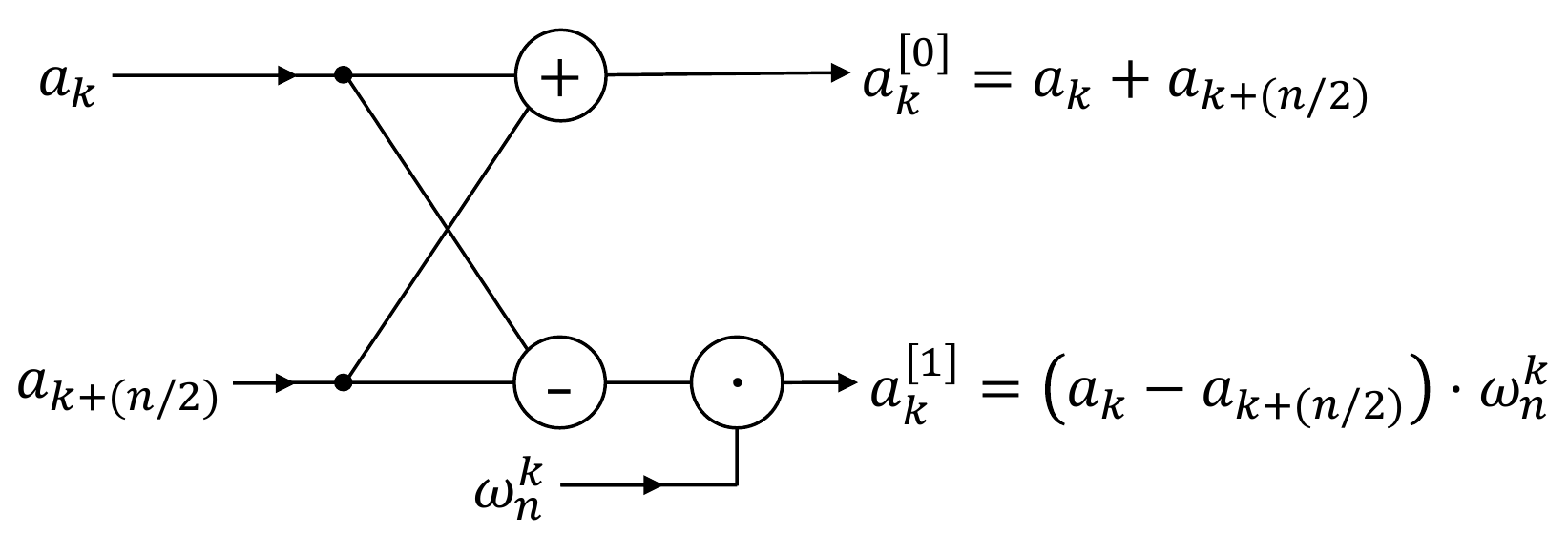}
\caption{Gentlemen-Sande butterfly}
\label{fig:gs-butterfly}
\end{minipage}
\end{minipage}
\hfill
\begin{minipage}[t]{.55\textwidth}\vspace{0pt}
\includegraphics[width=\columnwidth]{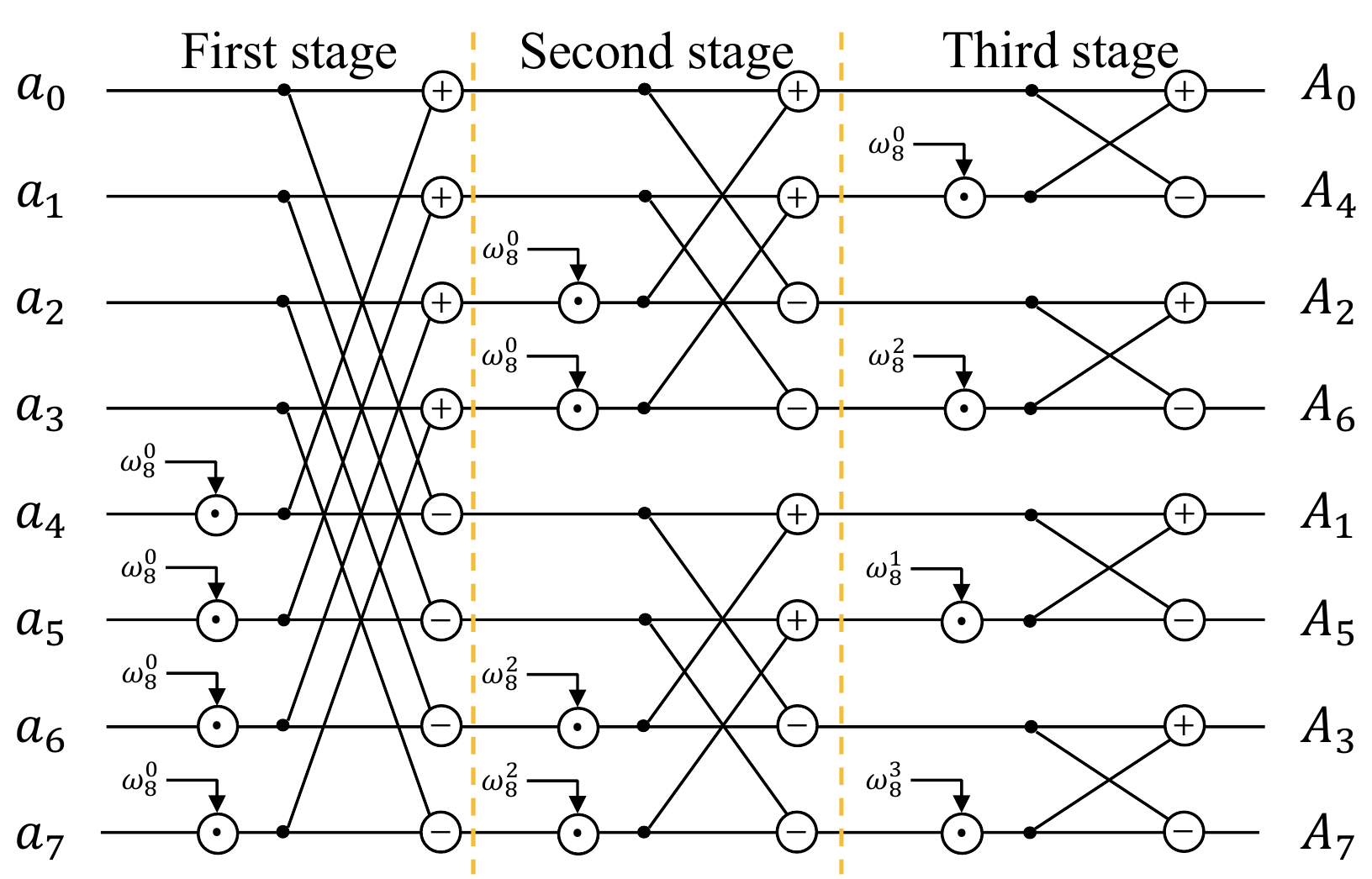}
\caption{Workflow of NTT/FFT with $n=8$ with CT butterfly}
\label{fig:8-point-butterfly}
\end{minipage}
\end{figure*}

Although FFT/NTT has lowered the computation complexity of polynomial multiplications to $O(n \log n)$, this operation still takes the majority of the time, reaching $70\%$ in some scenarios~\cite{ark}. Therefore, to further accelerate the FFT/NTT, people begin to design accelerators to further improve performance.

Most FFT/NTT implementations are constructed by iteratively executing a basic operation called butterfly. The most well-known butterfly strategies are Cooley-Tukey~(CT) butterfly~\cite{ct_butterfly}~(shown in Figure~\ref{fig:ct-butterfly}) and Gentlemen-Sande~(GS) butterfly~\cite{gs_butterfly}~(shown in Figure~\ref{fig:gs-butterfly}). These two strategies mainly consist of addition, subtraction and multiplication operations. CT butterfly is designed to execute the Equation \ref{eq:CT-butterfly-upperhalf} and \ref{eq:CT-butterfly-lowerhalf} in Appendix~\ref{app:ntt} by reusing the multiplication result. GS butterfly exploits a similar structure and can be used as the reverse operation of CT butterfly in applications. 

Based on the butterfly strategies, the FFT/NTT algorithm is divided into $\log n$ stages with $n / 2$ butterflies in each stage, as shown in Figure~\ref{fig:8-point-butterfly}, which presents an example of FFT with $n=8$. The input of the butterflies in FFT/NTT is the output of specific ones in the previous stage, which leads to a strict dependency chain from the first to the last stage. Data dependency among butterflies in adjacent stages continuously changes as the calculation proceeds. 

To accelerate FFT/NTT computation, the first attempt is to leverage data parallelism since data parallelism is widely supported by general hardware, such as CPU~(through SIMD instructions) and GPU~(through CUDA programming). However, data parallelism is only efficient when there are no dependencies among all pieces of data. Consequently, the data parallelism cannot be directly applied to FFT/NTT due to the strict data dependencies among adjacent butterfly operations. 

Another accelerating approach is to utilize pipeline parallelism, which can be implemented by designing customized hardware circuits via FPGA or ASIC. However, as Figure~\ref{fig:8-point-butterfly} indicates, to build a pipeline with large $n$, $(n \log n) /2$ separate butterflies should be implemented, which will consume too many hardware resources, causing either impossible circuits placement with FPGA or extremely expensive costs with ASIC. Moreover, data movement within such a large pipeline is almost impossible with state-of-the-art hardware technologies.

Therefore, naive data and pipeline parallelism are challenging. There are two possible solutions. First, it is possible to make FFT/NTT partly parallelizable based on the 4-step FFT/NTT algorithm~\cite{4-step-fft}. Readers may refer to the Appendix~\ref{app:4-step} for a detailed introduction to 4-step FFT/NTT. The second solution is inter-polynomial parallelism, executing operations over multiple polynomials concurrently with sufficient hardware resources. The scheme is called residue-polynomial-level parallelism~(rPLP), and we will discuss it in detail in \S\ref{sec:challenge_keyswitching}.


\subsubsection{Intensive Memory Access}
\label{sec:intense_memory}

Recent works have observed that overhead in memory access has become an essential bottleneck even if the accelerator is designed with modern hardware, such as FPGA or ASIC~\cite{f1, bts, craterlake, ark}. For example, let's consider a chip operating at $1$GHz and there're 40960 on-chip modular multiplication units. The chip is connected to the latest HBM3 with a bandwidth of $3$TB/s. Assume that the chip is executing CoeffToSlot with bootstrappable parameters ($n=2^{16}, L=23, {\rm dnum}=4$). All the calculation units are concurrently working and the bandwidth of the HBM3 is fully utilized. It only takes the chip $0.18$ms to finish all the multiplications, while it takes the HBM3 $2.1$ms~($11.7\times$ of computation time) to load the data used in CoeffToSlot~\cite{ark}.

The root cause of the problem is data inflation. Compared to plaintext computation, data size in polynomial operations largely increases for the following reasons.

In FFT/NTT, a group of pre-computed parameters called twiddle factors is mandatory for the calculations, \ie, $\omega_{n}^{k}$ in Figure~\ref{fig:ct-butterfly} and Figure~\ref{fig:gs-butterfly}. Because the twiddle factors can be reused throughout an entire FHE job, it is usually cached in the on-chip memory~(fast but extremely limited) of the accelerator to achieve high performance. For FFT, the number of twiddle factors is less than the number of input coefficients. However, for NTT, one group of twiddle factors is required for each modulus in RNS, which results in data inflation. In addition to the twiddle factors, the temporary results generated by each stage in FFT/NTT also place a high demand on the memory size.

As mentioned in~\S\ref{sec:ntt_complexity}, the 4-step FFT/NTT algorithm makes it possible to process a single FFT or NTT in parallel. Therefore, it is common to leverage this algorithm in recent designs. However, the 4-step FFT/NTT requires more pre-computed parameters than the original FFT/NTT algorithm (shown in Appendix~\ref{app:comp-4-step}). Although the algorithm greatly reduces the number of twiddle factors, it introduces $n$ additional pre-computed numbers called twisting factors for each RNS modulus~\cite{4-step-fft}. Therefore, the memory space requirement for pre-computed numbers is even more strict after applying the algorithm optimization.

Furthermore, due to the complex data dependency and large data size, read/write requests over the same block of memory from different butterflies are common. Such conflicts cause either large design complexity or severe performance degradation.


\subsubsection{Limited Generality}
\label{sec:limited_generality}
Degree $n$ of the polynomials are variant in different scenarios. For example, $n$ is related to security level and multiplicative depth.
However, it is challenging for these specific-designed hardware accelerators, such as FPGA and ASIC, to consistently provide optimal acceleration when $n$ varies. The reason is as follows. There are $(n\log n)/2$ butterflies in the entire FFT/NTT pipeline. When $n$ becomes larger, the consumption of hardware resources greatly increases. Furthermore, different $n$ leads to completely different data dependencies in FFT/NTT. Static circuit connections among butterflies cannot satisfy the requirement.


If we use a fixed $n$ to design the pipeline, the design suffers from resource underutilization or non-optimal performance. For example,
%
when an architecture designed for a certain degree is directly applied to accelerate FFT/NTT with a relatively larger degree, the calculation fails to achieve full pipelining because of insufficient multipliers. 
In contrast, adapting the design to scenarios where the degree is relatively smaller results in a tremendous waste of computing resources in the redundant stages.


\subsection{Challenges of Accelerating Key-switching}
\label{sec:challenge_keyswitching}



As mentioned in~\S\ref{sec:ct_maintenance}, key-switching is widely used for ciphertext conversion in homomorphic multiplication and permutation. The main challenge of accelerating key-switching is achieving generality, which we will demonstrate in detail in the following parts.

The key-switching algorithms used in state-of-the-art accelerators can be considered as different variants of the same algorithm, \ie, generalized key-switching\cite{homo-aes, better-bootstrapping}. Based on RNS decomposition, the modulus $Q$ is factorized into $L+1$ coprime moduli $q_0$, $q_1$, $\dots$, and $q_L$. In generalized key-switching, given a fixed integer parameter ${\rm dnum}\in [1, L+1]$ which stands for decomposition number, the moduli are grouped into ${\rm dnum}$ blocks and the partial production of moduli in each block is denoted as $Q_j=\prod_{i = j\alpha}^{(j+1)\alpha-1}q_i$, where $j\in[0, {\rm dnum})$ and $\alpha=\lfloor (L+1)/{\rm dnum}\rfloor $. Then the key-switching can be decomposed into calculations in each block, followed by the accumulation of results from different blocks. Each block's calculations are also called fast basis conversion, mainly consisting of multiplications and additions, while the major operation in accumulation is NTT. The selection of ${\rm dnum}$ greatly impacts the memory requirement and calculation complexity of key-switching. As ${\rm dnum}$ increases from $1$ to $L+1$, the overall memory consumption in key-switching and the workload for NTT strictly grows, while the workload for basis conversion decreases. The overall calculation workload also depends on the selection of other parameters, especially multiplicative depth $L$. Therefore, different $\rm dnum$s are suitable for various scenarios.

The variation of ${\rm dnum}$ poses generality challenges to the accelerators, especially for those with specific-designed circuits due to the high complexity of changing them. Since the proportion of different operations in the total workload is not static, the first challenge for the designers is to decide how to distribute the limited resources in the accelerator to different calculation units (\ie, NTT and basis conversion) to achieve balanced throughput for all the typical applications.

The second generality problem is how to decide the parallelism scheme. Since NTT and fast basis conversion require different schemes to be efficient, for the accelerators, how to parallel process polynomials is impacted by the relative workload of NTT and basis conversion, which is relevant to $\rm dnum$. The parallelism schemes adopted by existing works can be categorized into rPLP (residue-polynomial-level parallelism) and CLP (coefficient-level parallelism), two different polynomial access patterns for the accelerators to achieve high parallelism. Under RNS representation~(\S\ref{sec:rns}), we need to perform concurrent operations on multiple residue polynomials. With rPLP, each processor in the accelerator individually executes operations over a single residue polynomial. The parallelism is achieved by distributing multiple residue polynomials to different processors. With CLP, multiple processors in the accelerator collaboratively execute operations for a single residue polynomial, and $n$ coefficients in the same polynomial are distributed to different processors. There is no superior or inferior of these two schemes in terms of coefficient-wise operations like polynomial addition and dyadic multiplication. However, CLP introduces global data communication across different processors for NTT. Similarly, rPLP leads to extra data exchange in basis conversion. Therefore, both schemes may cause performance degradation in certain operations.

In key-switching, both basis conversion and NTT are dominant. The designer may compare the relative workload of basis conversion and NTT in the targeted applications to choose between CLP and rPLP. For example, when a relatively smaller ${\rm dnum}$ is chosen, which increases the importance of basis conversion, CLP is a preferred choice. Nevertheless, generality remains to be a challenge.

\subsection{Challenges of Accelerating Bootstrapping}
\label{sec:challenge_bootstraping}

In this section, we mainly analyze bootstrapping in word-wise FHE since it is more complicated than that in bit-wise FHE~(details in \S\ref{sec:ct_maintenance}), and is the focus of the latest FHE accelerators~\cite{craterlake,bts,ark}.
Bootstrapping in word-wise FHE consists of four complicated steps, making it the most time-consuming operation in word-wise FHE. The impressive overhead comes from the large consumption of multiplicative depth and enormous homomorphic permutations, which both lead to extremely high computation complexity and memory access.

\parab{Large Consumption of Multiplicative Depth:} As reported in~\cite{craterlake}, bootstrapping in the LSTM benchmark consumes $61\%$ of the maximum multiplicative depth~($35$ out of $57$ levels). The significant consumption of multiplicative depth mainly comes from the approximation schemes applied in bootstrapping, which include deep calculations.

Therefore, for FHE applications requiring unbounded multiplication depth, such as deep learning training, the maximum multiplicative depth $L$ must be large enough to support bootstrapping. Furthermore, since bootstrapping involves dramatic overhead, we should further increase depth $L$ to use more multiplicative depth to perform application computations before bootstrapping, thus reducing the frequency of bootstrapping during the whole calculation to guarantee the effectiveness of an application.

Consequently, the growth of depth $L$ leads to a larger modulus $Q$, and the degree of polynomial $n$ should be raised accordingly to reach a certain security level of RLWE~\cite{he-standard}. 
Large modulus and degree of polynomial lead to the growing size of ciphertexts and keys, thus significantly increasing the workload of polynomial calculations and memory consumption.

\parab{Enormous Homomorphic Permutations:} The second and fourth steps of bootstrapping, CoeffToSlot and SlotToCoeff, suffer from the high computational overhead of homomorphic operations, especially when the plaintext messages are densely packed in the ciphertext. In the implementation of~\cite{improved-fhe-boostrapping}, CoeffToSlot on a ciphertext, which packs 4096 plaintext messages, requires $\sim 30$ homomorphic permutations, leading to large computation complexity.
Moreover, homomorphic permutations introduce massive additional switching keys, dramatically inflating the memory space needed and increasing the bandwidth requirements.
In detail, the size of a single switching key is about ${\rm dnum} \times$ the size of a ciphertext, and permutations aiming at different rotation steps need distinct keys, which introduce ignorable memory storage and access overhead.
For example, according to the analysis in~\cite{ark}, with $n=2^{15}$, 40 different switching keys should be pre-computed and stored in preparation for permutations in the linear transformation of bootstrapping.
\section{FHE Accelerators}
\label{sec:fhe_acc_sok}

In this section, we will first comprehensively review the existing FHE accelerators in chronological order~(\S\ref{sec:sok-survey}). \bluehighlight{Then we briefly introduce how existing FHE software support these accelerators~(\S\ref{sec:sok-software}).} Finally, we will summarize their evolution trajectory to address particular challenges mentioned in the previous section, which establishes a qualitative connection among them~(\S\ref{sec:evo-sok}).

Table~\ref{tab:overview} provides an overview of FHE accelerators by demonstrating the hardware leveraged, the data parallelism schemes used, and the features of these FHE accelerators. We also make the following explanations to make it clear. First, the software/hardware co-design denotes whether the design leverages software's flexibility and hardware's efficiency to properly distribute and schedule the workload in the accelerator. In particular, we regard co-design is absent in all CPU-based and GPU-based works, which only include software design based on off-the-shelf hardware architectures. Second, similar to previous works~\cite{sok_fhe_compiler}, the FHE schemes are grouped into families of related schemes. For example, BFV stands for BFV~\cite{bfv}/BGV~\cite{bgv} and TFHE stands for TFHE~\cite{tfhe}/GSW~\cite{gsw}. Third, among the different features of the accelerators, we mainly focus on whether they are programmable and bootstrappable. Programmable refers to the accelerator's ability to support a variety of cryptographic parameters without hardware architecture reconfiguration. Bootstrappable means it is possible to execute bootstrapping and achieve unlimited FHE operations with the accelerator. The half circle~(\halfcirc[0.6ex]) here represents that the accelerator supports bootstrapping, but the performance is far from meeting practical requirements.

Worth noting, although the FHE accelerators rely on the acceleration of polynomial computations, such as NTT, it is not reasonable to categorize works that only accelerate these basic modules as FHE accelerators. For example, FFT accelerators are commonly applied in other fields, such as signal processing, but most of them cannot be directly used to boost the performance of FHE schemes. Our survey will not cover them but focus on accelerators that explicitly improve the efficiency of FHE schemes.

\begin{table}[t]
\footnotesize
\centering
\begin{tabularx}{\textwidth}{l c c c c c c c c c}
\toprule
\multirow{2}{*}{\bf Name} 
& \multirow{2}{*}{\shortstack[c]{\bf Hardware \\ \bf Target}}
& \multirow{2}{*}{\shortstack[c]{\bf Software/Hardware \\ \bf Co-design}}
& \multirow{2}{*}{\shortstack[c]{\bf Polynomial \\ \bf Parallelism}}
& \multicolumn{3}{c}{\bf Supported Schemes} 
& \multicolumn{2}{c}{\bf Supported Features} 
& \multirow{2}{*}{\shortstack[l]{\bf Open- \\ \bf source}}\\
& & & & 
BFV & CKKS & TFHE 
& Programmable & Bootstrappable \\
\midrule
cuHE~\cite{cuhe-url} & GPU & \emptycirc[0.6ex] & rPLP & \emptycirc[0.6ex] & \emptycirc[0.6ex] & \fullcirc[0.6ex] & \emptycirc[0.6ex] & \fullcirc[0.6ex] & \fullcirc[0.6ex] \\
cuFHE~\cite{cufhe-url} & GPU & \emptycirc[0.6ex] & rPLP & \emptycirc[0.6ex] & \emptycirc[0.6ex] & \fullcirc[0.6ex] & \emptycirc[0.6ex] & \fullcirc[0.6ex] & \fullcirc[0.6ex] \\
nuFHE~\cite{nufhe-url} & GPU & \emptycirc[0.6ex] & rPLP & \emptycirc[0.6ex] & \emptycirc[0.6ex] & \fullcirc[0.6ex] & \emptycirc[0.6ex] & \fullcirc[0.6ex] & \fullcirc[0.6ex] \\
HEAT~\cite{heaws} & FPGA & \fullcirc[0.6ex] & rPLP & \fullcirc[0.6ex] & \emptycirc[0.6ex] & \emptycirc[0.6ex] & \emptycirc[0.6ex] & \emptycirc[0.6ex] & \fullcirc[0.6ex] \\
HEAX~\cite{heax} & FPGA & \emptycirc[0.6ex] & rPLP & \emptycirc[0.6ex] & \fullcirc[0.6ex] & \emptycirc[0.6ex] & \emptycirc[0.6ex] & \emptycirc[0.6ex] & \emptycirc[0.6ex] \\
HEXL~\cite{hexl} & CPU & \emptycirc[0.6ex] & rPLP & \fullcirc[0.6ex] & \fullcirc[0.6ex] & \emptycirc[0.6ex] & \fullcirc[0.6ex] & \emptycirc[0.6ex] & \fullcirc[0.6ex] \\
HEXL-FPGA~\cite{hexl-fpga-url} & FPGA & \emptycirc[0.6ex] & rPLP & \fullcirc[0.6ex] & \fullcirc[0.6ex] & \emptycirc[0.6ex] & \fullcirc[0.6ex] & \emptycirc[0.6ex] & \fullcirc[0.6ex] \\
100$\times~$~\cite{over100} & GPU & \emptycirc[0.6ex] & rPLP + CLP & \fullcirc[0.6ex] & \fullcirc[0.6ex] & \emptycirc[0.6ex] & \fullcirc[0.6ex] & \fullcirc[0.6ex] & \fullcirc[0.6ex] \\
F1~\cite{f1} & ASIC & \fullcirc[0.6ex] & rPLP & \fullcirc[0.6ex] & \fullcirc[0.6ex] & \emptycirc[0.6ex] & \fullcirc[0.6ex] & \halfcirc[0.6ex] & \emptycirc[0.6ex] \\
CraterLake~\cite{craterlake} & ASIC & \fullcirc[0.6ex] & CLP & \fullcirc[0.6ex] & \fullcirc[0.6ex] & \emptycirc[0.6ex] & \fullcirc[0.6ex] & \fullcirc[0.6ex] & \emptycirc[0.6ex] \\
BTS~\cite{bts} & ASIC & \fullcirc[0.6ex] & CLP & \fullcirc[0.6ex] & \fullcirc[0.6ex] & \emptycirc[0.6ex] & \fullcirc[0.6ex] & \fullcirc[0.6ex] & \emptycirc[0.6ex] \\
ARK~\cite{ark} & ASIC & \fullcirc[0.6ex] & rPLP + CLP & \fullcirc[0.6ex] & \fullcirc[0.6ex] & \emptycirc[0.6ex] & \fullcirc[0.6ex] & \fullcirc[0.6ex] & \emptycirc[0.6ex] \\
Poseidon~\cite{poseidon} & FPGA & \fullcirc[0.6ex] & CLP & \fullcirc[0.6ex] & \fullcirc[0.6ex] & \emptycirc[0.6ex] & \fullcirc[0.6ex] & \fullcirc[0.6ex] & \emptycirc[0.6ex]\\
FAB~\cite{fab} & FPGA & \fullcirc[0.6ex] & CLP & \fullcirc[0.6ex] & \fullcirc[0.6ex] & \emptycirc[0.6ex] & \emptycirc[0.6ex] & \fullcirc[0.6ex] & \emptycirc[0.6ex] \\
TensorFHE~\cite{tensorfhe} & GPU & \fullcirc[0.6ex] & rPLP & \fullcirc[0.6ex] & \fullcirc[0.6ex] & \emptycirc[0.6ex] & \fullcirc[0.6ex] & \fullcirc[0.6ex] & \emptycirc[0.6ex] \\
\bottomrule
\end{tabularx}
\caption{Overview of all FHE accelerators~(in chronological order).}
\label{tab:overview}
\end{table}

\subsection{Survey of Existing FHE Accelerators}
\label{sec:sok-survey}
\bluehighlight{For all the FHE accelerators, the common building block is the FFT/NTT module, as it is the bottleneck of polynomial calculations and directly determines the performance. We need to point out that most existing accelerators~(except for the GPU-based design nuFHE~\cite{nufhe-url}) do not implement both FFT and NTT computations since FFT contains floating-point operations, while NTT only includes integer operations. They require completely different hardware circuits for acceleration. Therefore, to simplify the design complexity and reduce the hardware resource overhead, the accelerator designers usually accelerate one of FFT and NTT, and therefore can only support FHE algorithms using the corresponding operations, as we listed in Table~\ref{tab:overview}.

By connecting the FFT/NTT module with other computation units with optimized data paths, the accelerators can finish their target FHE jobs. Different works have variant designs for FFT/NTT modules and data paths, targeting their acceleration goals. For earlier works, since their computation scenarios are limited, they usually have fixed designs for NTT/FFT computation units and fixed calculation workflow to reduce design complexity. As the application scenarios continue to increase, later works tend to design programmable NTT/FFT units and data paths to support different parameter settings and various applications. Moreover, later works find it is important to introduce more on-chip memory spaces with customized devices, as the memory of general hardware like GPU and FPGA becomes insufficient because the FHE applications become more complicated and require more calculations. As a result, Application-Specific Integrated Circuit~(ASIC) becomes a preferred choice for designers who can afford it.}

\subsubsection{cuHE/cuFHE}
CUDA Homomorphic Encryption Library~(cuHE) was proposed by Dai \etal in 2015~\cite{cuhe,cuhe-url}.
cuHE uses GPU for acceleration and provides CUDA implementation of NTT and CRT. 
CUDA-accelerated Fully Homomorphic Encryption Library~(cuFHE) was proposed by Dai \etal in 2018~\cite{cufhe-url}.
cuFHE leverages implementations of cuHE to boost the performance of TFHE~\cite{tfhe}.
Both cuHE and cuFHE are open-sourced standalone acceleration libraries~\cite{cuhe-url,cufhe-url} and do not provide official integration with FHE libraries.
However, they only support accelerating polynomial operations with NTT, while an FFT-based implementation may achieve higher performance for TFHE. Moreover, besides limited functions, cuFHE adopts fixed cryptographic parameters, which cannot be configured to 
achieve high generality.

\subsubsection{nuFHE}
GPU-powered Torus FHE implementation~(nuFHE) was launched by NuCypher in 2018~\cite{nufhe-url}. 
Similar to cuHE/cuFHE, nuFHE also adopts GPU to accelerate FHE schemes. 
However, different from them, nuFHE provides either FFT or NTT to improve performance.
nuFHE is an open-source standalone library~\cite{nufhe-url} and provides Python APIs.
With FFT, nuFHE achieves better performance compared with cuFHE. But it shares the similar disadvantage of 
limited generality.

\subsubsection{HEAT}
HEAT was proposed by Roy \etal in 2019~\cite{heaws}.
Different from cuHE, cuFHE, nuFHE, HEAT targets accelerating a word-wise FHE scheme: BFV. Moreover, besides general hardware which has been used in previous works, HEAT further leverages FPGA to achieve a more flexible hardware design. Thus, HEAT utilized a heterogeneous ARM-FPGA co-designed architecture implemented on Xilinx ZCU102 Evaluation Kit~\cite{zcu102}.
The authors further migrated HEAT to the f1 instance of Amazon AWS in 2020~\cite{heaws_aws}. Both implementations are open-sourced. The implementation of HEAT on ZCU102 comprises several hardware coprocessors on FPGA and a multi-core~(4 cores) ARM processor. The parallel coprocessors can efficiently execute primitive operations in BFV, including addition, subtraction, multiplication, modulus switching, and NTT. The ARM cores control the coprocessors' workflow and manage the network connection to the applications. 
By using the software as a workflow controller, HEAT can perform polynomial arithmetics and homomorphic operations like key-switching in BFV by combining different primitive operation units.

Considering the limited on-chip memory~(BRAM/URAM) and calculation units~(DSP), it is impossible to implement a fully pipelined NTT processor on FPGA. Instead, the authors instantiated two CT butterfly units for each NTT core and accomplished the NTT by reusing them. This design is a compromise solution due to limited resources, thus inevitably leading to non-optimal performance. Another problem caused by resource constraints is the relatively smaller polynomial degree. The polynomial degree supported by HEAT is $4096$, which does not fulfill most practical requirements and significantly limits its generality. 

\subsubsection{HEAX}
HEAX was proposed by Microsoft in 2020~\cite{heax}. It provides a highly performant hardware architecture to accelerate the operations of CKKS. HEAX is not an open-source project. The acceleration foundations of HEAX are the primitive modules: NTT and multiplication modules. Each module consists of multiple calculation cores, which could be adjusted to match the required throughput.

Different from HEAT, which depends on the software to manipulate hardware primitive operations~(\eg, NTT and multiplication) when implementing key-switching, HEAX implements a specific key-switching module to minimize the overhead of software and hardware interaction. Following the workflow of key-switching, the key-switching module instantiates multiple primitive modules and BRAMs to construct the pipeline. The modules can be adjusted to balance the throughput in the pipeline based on specific cryptographic parameters.

However, the adjustment of HEAX is accomplished by physically modifying the modules. The FPGA has to be reconfigured to adapt to different cryptographic parameters, which limits the generality of the design.


\subsubsection{HEXL}
Unlike previous works relying on specific hardware devices, such as GPU and FPGA, Intel proposed Intel Homomorphic Encryption Acceleration Library~(HEXL) in 2021~\cite{hexl}. HEXL exploits the SIMD features provided by Intel CPUs, which are easily accessible, to provide plug-and-play acceleration capacities for FHE. HEXL has been integrated with PALISADE~\cite{palisade-url}, Microsoft SEAL~\cite{seal-url} and HElib~\cite{helib-url} by replacing the underlying arithmetic implementations. It is open-sourced, and the code is hosted on Github~\cite{hexl-url}.

With the single-threaded implementation of primitive operations, including NTT/iNTT and dyadic multiplication, based on the Intel AVX-512/AVX512IFMA instructions~\cite{avx512-url}, HEXL reaches high single-core acceleration. Since HEXL is thread-safe, users can achieve better performance by paralleling different operations with multi-threading. 
However, due to architecture deficiency, the overall performance largely degrades when there are many threads. For example, too many threads with SIMD cause significant heat dissipation. As a result, core frequency is dramatically reduced when reaching the TDP~\cite{simd_performance}. Moreover, similar to the CPU-based acceleration libraries in other domains, HEXL also suffers from slow memory access, further reducing its overall performance.

\subsubsection{HEXL-FPGA}
\label{sec:hexl-fpga}
To compensate for the disadvantages of HEXL, Intel further proposed HEXL-FPGA in 2021~\cite{hexl-fpga-url}.
HEXL-FPGA offers the HLS-based~(high level synthesis) implementation of NTT/iNTT, multiplication, and key-switching. Since HEXL-FPGA is open-sourced, users can compile each of the functions into an individual bitstream and program the FPGA device. HEXL-FPGA can be integrated with aforementioned HEXL to accelerate corresponding FHE libraries.
HEXL-FPGA is open-sourced and under active development~\cite{hexl-fpga-url}.

However, there are inherent problems with HLS-based solutions. HLS designs are written in high-level language and rely on the compiler to convert the codes into hardware design. Due to the essential difference between the hardware and software, the compiling process may lead to redundant resource consumption and complex workflow synchronization~\cite{hls}, thus leading to suboptimal performance. Therefore, it's hard for HEXL-FPGA to fully utilize the advantages of FPGA. Moreover, HEXL-FPGA only supports a limited range of cryptographic parameters, which will be demonstrated in \S\ref{sec:eval-results}.

\newcommand{\ckksgpu}{100$\times$\xspace}

\subsubsection{\ckksgpu}
\ckksgpu was proposed by Jung \etal in 2021\cite{over100}. It provides higher acceleration of CKKS than previous works (\ie, HEXL-FPGA and HEAX) with the powerful V100 GPU~\cite{v100}. The reason is that V100 has more hardware resources due to better semiconductor manufacturing processes and works at a much higher frequency than the FPGAs adopted in previous works. \ckksgpu introduces memory-centric optimizations to increase end-to-end performance and achieves over 100$\times$ acceleration ratio compared to the single-thread CPU implementation.

The implementation of NTT in \ckksgpu is built based on the hierarchical approach in~\cite{gpuntt}. Its basic idea follows the 4-step FFT/NTT algorithm~\cite{4-step-fft}. The implementation of NTT is divided into two separate kernels, making it possible to cache all input data in the shared memory. Moreover, due to the limited size of registers in GPU, both kernels further decompose the NTT based on a generalized version of the 4-step FFT/NTT algorithm. In addition to the decomposition, schemes including coalesced memory access and on-the-fly twiddle factors generation are leveraged to decrease the overhead in memory access.

However, although the memory accesses in the basic operations have been optimized, the end-to-end performance of a FHE task is still bottlenecked by the bandwidth of the main memory.
To relieve the bottleneck, the authors of \ckksgpu tend to fuse multiple kernels into a single kernel, which allows the data cached on the chip to be reused by a series of consecutive operations and reduces global memory accesses.
Nevertheless, since GPU's architecture is designed for calculations between small numbers (\eg, FP16), it does not offer large on-chip memory. As a result, the intensive memory access still degrades the performance of \ckksgpu.


\subsubsection{F1}
\label{sec:f1}
To solve the problems of insufficient resources~(\eg, the FPGA-based ones) and unsuitable fixed architecture~(\eg, the CPU or GPU-based ones), recent works have shifted to explore the potential of application-specific integrated circuits~(ASICs).
Following this trend, F1 was proposed by Feldmann \etal in 2021~\cite{f1}.
To pursue more practical acceleration for FHE schemes, F1 is the first programmable FHE accelerator with a dedicated architecture, which denotes it can support several FHE schemes with a large range of cryptographic parameters. F1 is not open-sourced. 

At its core, F1 implements $16$ computation clusters, each containing several primitive function units~(FU), including NTT, modular multiplication, modular addition, and automorphism. Different units can compose high-level FHE operations such as key-switching. All the FUs are pipelined and vectorized to process $128$ elements in each cycle. Therefore, polynomials with degrees that are multiples of $128$ can be handled by sequentially feeding the operands to the pipeline. Specifically, to implement NTT with the $128$-element processor, F1 leverages the 4-step FFT/NTT algorithm~\cite{4-step-fft} to decompose a NTT into multiple vectorized operations with much fewer input elements. 
F1 tries to minimize data movement overhead by proposing a hierarchical storage system. The off-chip high-bandwidth memory~(HBM) is the global memory that directly interacts with the CPU server. A $64$MB scratchpad built on $16$ banks of SRAM is designed as the on-chip cache and fetches data from HBM. The computation clusters communicate with the scratchpad and store the data used for the current operation in the limited vector registers. The communication between the scratchpad and clusters depends on a complex fully connected network~(three $16\times 16$ crossbars).

The design of F1 has three main problems. First, the performance of F1 highly relies on sufficient parallelism among 16 clusters and efficient data movement in the accelerator, which places a high demand on the software compiler for operation scheduling. Considering that FHE computations vary significantly in terms of different workflows and different parameter settings, the compiler is complicated. However, F1 does not provide many details about its compiler. The second problem is that F1 uses a fixed key-switching algorithm~(generalized key-switching with ${\rm dnum} = L+1$). If the multiplicative depth $L$ of ciphertexts is high, key-switching is extremely slow. Last, F1 only supports non-packed bootstrapping~\cite{ckks-bootstrapping}, which only works for ciphertext with a single number packed in the polynomial and is far from practical in real-world applications. The reason is that the maximum degree of polynomial that F1 supports via the 4-step FFT/NTT algorithm is $16384$, which is too small for F1 to execute fully packed bootstrapping~\cite{efficient-bootstrapping} under $80$-bit or $128$-bit security level of RLWE. Without efficient bootstrapping, F1 struggles to evaluate deep arithmetic functions.

\subsubsection{CraterLake}
CraterLake was proposed by Samardzic \etal in 2022~\cite{craterlake}. It is not an open-source project. CraterLake is a follow-up to F1 and targets unbounded-depth homomorphic multiplications. Consequently, CraterLake uses the fully packed bootstrapping~\cite{efficient-bootstrapping} to refresh the multiplicative depth of ciphertexts. To balance the frequency of bootstrapping and the size of ciphertexts, the authors of CraterLake chose the number of multipliers required per homomorphic multiplication as the criterion to evaluate the overall computational complexity of an FHE program. 
Moreover, the authors followed the idea of the vectorized unit in F1 and logically designed the whole accelerator as a single $2048$-element vectorized processor to handle the large ciphertext, which requires the maximum degree of polynomial and multiplicative depth to be $65536$ and $60$ respectively.
However, such a design can not be naively implemented on ASIC considering the complex combinational and sequential logic circuits. Thus, in CraterLake, the $2048$-element processor is physically composed of eight $256$-element groups.

Similar to F1~\cite{f1}, CraterLake implements several functional units~(FU) in each $256$-element group. The difference is the two additional FUs in CraterLake, \ie, Change-RNS-base~(CRB) and switching key generator~(KSHGen), and both units are introduced to further optimize key-switching.
The CRB unit, which mainly executes modular multiplication and modular summation, is specifically designed to accelerate fast basis conversion when $\rm dnum$ is relatively small (\eg, $\rm dnum=1$). To save additional memory space, the KSHGen unit generates switching keys on the fly, which were pre-computed and cached in the device memory in previous works, such as F1~\cite{f1} and HEAX~\cite{heax}.

Another major difference between CraterLake and previous works is that CraterLake adopts CLP as the data parallelism scheme rather than rPLP due to the increasing workload of fast basis conversion. With the CLP scheme, CraterLake only processes one polynomial operation simultaneously, and the coefficients of polynomials are distributed to the $256$-element groups with a static distribution strategy, leading to three significant advantages. First, as a particular coefficient can only be assigned to a specific group, the complex $16\times 16$ crossbar in F1, which deals with the dynamic data exchange between on-chip scratchpad and groups, is no longer required. Data movement caused by operations among different polynomials can be eliminated as well. Second, the parallelism of CraterLake is not influenced by the varying multiplicative depth or numbers of concurrent operations, which maintains the performance throughout an entire FHE program and simplifies the software scheduler.

Although CraterLake is programmable enough to support different cryptographic parameters, the FUs, especially NTT units, may face functionality limitations or resource underutilization because of the static pipelined circuit, as we mentioned in~\S\ref{sec:limited_generality}. The computation resources in CraterLake can be fully utilized if and only if $n=65536$, which is a common problem for works containing a pipelined NTT circuit~(\eg, F1 and ARK).
Besides, in CraterLake, the CRB unit occupies $34\%$ on-chip area, reducing the performance of other operations, like NTT. Therefore, in the scenarios where larger $\rm dnum$ is optimal, the CRB units are underutilized and CraterLake cannot deliver sufficient acceleration.

\subsubsection{BTS}
BTS was proposed by Kim \etal in 2022~\cite{bts}. It is a follow-up to \ckksgpu and shares similar design goals with CraterLake, \ie, a bootstrappable and programmable hardware architecture for word-wise FHE. 

Because of the similar goal, many optimization approaches in BTS are close to those in CraterLake, including the basis conversion unit, CLP parallelism scheme and on-the-fly data generation.
The most significant difference between CraterLake and BTS is the pattern in which they place basic functional units, thus leading to different implementations of basic operations. The architecture of BTS is close to that of modern GPUs, consisting of a $64\times 32$ two-dimensional (2D) array of processing elements~(PE). Each PE comprises a basis conversion unit and a $2$-point NTT/iNTT unit (\ie, a single butterfly unit). The $2048$ PEs are interconnected via vertical and horizontal crossbars. An example of $131072$-point NTT with the architecture works as follows.
First, each PE performs a $64$-point NTT independently.
Second, $64$ PEs in the same row perform a $64$-point NTT with data synchronization through horizontal crossbars.
Third, $32$ PEs in the same column perform a $32$-point NTT with data synchronization through vertical crossbars.
This process generalized the 4-step FFT/NTT algorithm. Each PE plays a similar role to the GPU threads, and the crossbars accomplish thread synchronization.
To facilitate the unique computational pattern, the authors placed multiple smaller local scratchpads in each PE instead of putting a large global one that communicates with all the PEs. 
The local scratchpad is directly connected with HBM and eliminates additional overhead from the hierarchical data distribution.

However, a potential drawback of BTS is that the 2D array structure cannot execute NTT with a fully pipelined workflow. As mentioned before, NTT operation in BTS consists of three sequential steps and all the PEs are involved in each step. Therefore, no pipeline parallelism can be achieved if multiple NTT operations should be processed, which may decrease the end-to-end performance. Similar to F1 and \bluehighlight{CraterLake}, BTS is not open-sourced.


\subsubsection{ARK}
ARK was proposed by Kim \etal in 2022~\cite{ark}. It consists of four vectorized processing clusters, similar to the $256$-element groups of CraterLake. Each cluster contains several primary operation units. Different from previous works that target accelerating bootstrapping only from hardware angle~\cite{craterlake,bts}, ARK is the first work that analyzes and modifies the bootstrapping algorithm, thus achieving an algorithm and hardware co-design to optimize the accelerator's performance. Specifically, the authors proposed multi-hop homomorphic rotation and on-the-ﬂy residue extension to significantly reduce the size of switching keys and plaintexts used in bootstrapping, respectively, which lowers the difficulties in hardware design.

As mentioned earlier in this section, the workload of basis conversion is heavy in bootstrapping, making CLP a preferred parallelism scheme for bootstrappable accelerators like CraterLake and BTS. But CLP also introduces communication overhead in NTT. According to ARK, NTT and basis conversion separately take up $54.8\%$ and $34.2\%$ of the overall computational workload given practical bootstrappable parameters. Since both functions are crucial to end-to-end performance, ARK leverages both schemes, \ie, CLP for basis conversion and rPLP for the other operations including NTT. A switching mechanism between the two data distribution patterns is also used by ARK, especially for functions that contain both NTT and basis conversion, such as key-switching.
In ARK's paper, this hybrid parallelism is proven more efficient than rPLP, but whether it is better than CLP is not discussed.

\subsubsection{Poseidon}
Poseidon was proposed by Yang \etal in 2023 to design a FPGA-based FHE accelerator that could support bootstrapping~\cite{poseidon}. Compared to previous FPGA-based accelerators, Poseidon leverages several optimization operations, such as radix-based NTT, optimized automorphism, \etc, to improve the resource efficiency of its implementation. Furthermore, Poseidon utilizes a more modern FPGA, U280~\cite{u280}, with high-bandwidth memory~(HBM)~\cite{hbm} support. Therefore, Poseidon manages to implement the complicated bootstrapping algorithm on FPGA. However, Poseidon should still suffer the limitations of aforementioned FPGA-based solutions.

\subsubsection{FAB}
FAB was proposed by Agrawal \etal in 2023~\cite{fab}. Similar to Poseidon~\cite{poseidon}, FAB is a FPGA-based FHE accelerator that can support bootstrapping~\cite{fab}. FAB optimizes the on-chip memory access based on the workflow of bootstrapping algorithm, to eliminate the bottleneck caused by memory access. Similar to Poseidon~\cite{poseidon}, FAB also uses U280~\cite{u280} FPGA with HBM support.

\subsubsection{TensorFHE}
TensorFHE was proposed by Fan \etal in 2023. By transforming NTT into multiple sequential general matrix multiplications, TensorFHE can use the Tensor Core Units~(TCU)~\cite{tcu} in modern GPUs to accelerate the NTT computation~\cite{tensorfhe}. The core idea of TensorFHE is to leverage fine-grained operation batching to utilize the data parallelism features provided by a GPU. With state-of-the-art GPUs, such as V100~\cite{v100} and A100~\cite{a100}, the authors claim that TensorFHE can outperform other accelerators implemented with CPU, GPU and FPGA. TensorFHE could achieve comparable performance as F1~\cite{f1} in some scenarios, due to the latest semi-conduct technologies adopted by these modern GPUs. TensorFHE should still suffer from the aforementioned non-optimal acceleration due to GPU's architectural deficiency.

\bluehighlight{
\subsection{Upper-layer Software Support}
\label{sec:sok-software}
To effectively utilize the accelerators mentioned above, proper upper-layer software support is crucial. However, integrating FHE software with these accelerators poses significant challenges. One key obstacle is the need to establish a well-defined and unified interface that accommodates different FHE accelerators, each utilizing distinct data structures for input/output and requiring diverse workflows. Addressing this challenge is signifant to the research community, as achieving transparency of FHE accelerators to upper-layer software is essential for practical real-world deployment.

A promising advancement towards solving this problem is the OpenFHE project~\cite{openfhe}. OpenFHE is an open-source project that offers efficient and extensible implementations of the latest FHE schemes. Notably, OpenFHE introduces a hardware abstraction layer~(HAL) designed to support CPU, GPU, FPGA, and ASIC-based FHE accelerators. The HAL provides C++ abstract classes that enable accelerators to implement their own backend logic. These classes encompass essential mathematical operations (such as NTT, iNTT, vector addition, \etc.) and lattice/polynomial layer functionalities (including RNS subroutines). It is worth mentioning that OpenFHE already includes a hardware backend implementation for HEXL~\cite{openfhe-hexl,hexl}.

By addressing the interface and compatibility challenges and offering a flexible HAL, OpenFHE empowers researchers and developers in the FHE community to seamlessly integrate diverse FHE accelerators into their upper-layer software solutions. We believe that this advancement will shed light on bridging the gap between FHE applications and accelerators.
}

\subsection{Observations on Evolution of Existing Works}
\label{sec:evo-sok}

Based on the above survey, we make the following observations on the evolution of existing FHE accelerators and the underlying connection among them. We also illustrate how they address particular challenges mentioned in \S\ref{sec:fhe_inefficiency}.

\parab{From General Hardware to Application-Specific Integrated Circuit:} Most latest FHE accelerators choose Application-Specific Integrated Circuit~(ASIC) as their hardware platform for two reasons. First, general-purpose hardware, such as CPU and GPU, suffer from fixed architecture and limited on-chip memory, which fail to address the challenges mentioned in \S\ref{sec:fhe_inefficiency}. Second, FPGA suffers from relatively limited programmable resources and low operational frequency, which restricts it from reaching high performance. Although some latest FHE accelerators still utilize FPGA or GPU in their design~\cite{poseidon,fab,tensorfhe}, they still suffer the mentioned limitations, thus leading to suboptimal performance.

The advantage of applying ASICs to FHE accelerators is that designers can optimize the hardware architecture with high flexibility and utilize state-of-the-art technologies of the computer architecture community according to the requirement of FHE computations. Recent ASIC-based works adopt specially designed architecture, such as crossbars, to enable complex polynomial operations described in~\S\ref{sec:ntt_complexity}. They also use high-speed memory, including HBM and SRAM, to accelerate the intensive memory access mentioned in~\S\ref{sec:intense_memory} and~\S\ref{sec:challenge_bootstraping}. While promising, this design choice also introduces several problems: 1) most modern ASIC-based FHE accelerators are expensive to produce, especially when the chip size is up to hundreds of square millimeters, \eg, a 14nm ASIC of 100mm$^2$ requires millions of US dollars to tape out; 2) most of these ASIC-based solutions are not open-sourced due to reasons such as IP restrictions, making them difficult to reproduce, thus undesirable for the research community.

\parab{Software/Hardware Co-design:}
Early FHE accelerators 
perform certain simple operations, such as NTT, after receiving a single instruction from upper-layer software without any instruction scheduling~\cite{heaws,heax}. Since these operations follow the static hardware workflow, which cannot be dynamically adapted, the accelerators suffer from limited generality~(\S\ref{sec:limited_generality}). Moreover, when accelerating an end-to-end application containing massive high-level operations such as bootstrapping~(\S\ref{sec:challenge_bootstraping}), the overhead of frequent interactions between software and hardware cannot be ignored.

To overcome these problems, recent FHE accelerators adopt software/hardware co-designed approaches. Specifically, they first leverage hardware-friendly algorithms to divide the hardware resources into multiple computation clusters which can be independently scheduled~\cite{f1,craterlake,bts,ark}. Furthermore, the designers can make extensive use of the software's flexibility to apply adaptations according to the specific application to efficiently support complicated end-to-end applications.

However, the current software design of FHE accelerators fails to leverage knowledge from some of the latest techniques, \eg, various FHE compilers~\cite{eva,e3}. We identify it as a potential future direction in \S\ref{soft-hardware-co-design}.


\parab{Enhanced Programmability:}
Programmability, \ie, supporting different cryptographic parameters without hardware architecture reconfiguration, is not achieved in most of the early works due to the following two reasons. First, early works rarely focus on end-to-end acceleration for real-world applications. Thus, they do not have such a need. Second, programmability is not easy to achieve, considering the complexity of FFT/NTT. 

Recent FHE accelerators have begun to adopt the 4-step FFT/NTT algorithm not only to increase parallelism but also to improve the architecture's programmability to handle different parameters, specifically various $n$. Readers may refer to Appendix~\ref{app:comp-4-step} for more details. The improved programmability alleviates the generality limitation in polynomial computations described in~\S\ref{sec:limited_generality}, although the problems are not eradicated. Currently, an unresolved issue is the generality challenge in key-switching~(\S\ref{sec:challenge_keyswitching}). Recent works focusing on deep calculations show unoptimized performance in shallow tasks. As described in~\cite{craterlake}, in shallow benchmarks without bootstrapping~($L$ is between 4 and 8), CaterLake is slower than F1 due to the underutilization of the basis conversion units that occupy a large proportion of hardware resources.

\parab{Unlimited Depth of Operations:}
Interestingly, the latest few works~(\ie, CraterLake~\cite{craterlake}, BTS~\cite{bts}, and ARK~\cite{ark}) show a similar tendency of accelerating bootstrapping to achieve unlimited 'fully' homomorphic operations. This common goal leads to the convergence of multiple design choices. First, optimized algorithms, including 4-step FFT/NTT and generalized key-switching, are preferred because of their advantages in handling large ciphertexts. Recent studies are willing to allocate a considerable amount of hardware resources to corresponding structures such as global transpose and fast basis conversion. Second, recent works invest much effort in eliminating the memory access bottleneck. They tend to reduce the memory overhead at the cost of introducing extra calculations, like on-the-fly generation of essential parameters. Last but not least, since all these works have emphasized the importance of basis conversion, CLP is becoming the primary parallelism scheme, contrasting with rPLP in the earlier designs.
\section{Evaluation}
\label{sec:evaluation}

\subsection{Evaluation Methodology}
\label{sec:eval-methodology}
In this section, we provide a quantitative comparison of these existing FHE accelerators.
As introduced in \S\ref{sec:fhe_acc_sok}, some FHE accelerators are open-sourced. We will use our testbed to reproduce the results of some representative ones. Since these open-sourced FHE accelerators do not provide direct support for end-to-end algorithms, we mainly evaluate how they accelerate the performance of NTT and key-switching. In this paper, we evaluate HEXL, HEXL-FPGA, and \ckksgpu as word-wise FHE schemes. We also evaluate two bit-wise FHE accelerators: cuFHE and nuFHE.

Second, since some latest FHE accelerators are not open-sourced, we will directly use the results from their original papers. For example, the paper of HEAX and F1 provides its performance of accelerating NTT and key-switching while TensorFHE also provides its performance of NTT, thus we will align these results with our testbed results. Moreover, some latest accelerators, such as F1, BTS, ARK, Poseidon, FAB and TensorFHE provide end-to-end performance results \bluehighlight{for accelerating real-world applications, we will also include these results in our paper. Specifically, we demonstrate the performance of two deep applications, \ie, ResNet20~\cite{resnet20} and Logistic Regression~\cite{lr}, and two shallow applications, \ie, LoLa-MNIST and LoLa-CIFAR~\cite{lola}. Deep applications contain many ciphertext multiplications and therefore require bootstrapping, while shallow applications only include limited multiplications and do not need bootstrapping. For the detailed settings of the end-to-end tests, we refer the readers to the original papers of the accelerators.}


\parab{Testbed settings:} We use a single X86 server as our testbed. The server is equipped with an Intel(R) Xeon(R) Gold 5115 CPU running at $2.40$GHz and $128$GB RAM. The CPU supports AVX-512 FMA~\cite{avx512-url}. \bluehighlight{The operating system is Ubuntu 18.04.5 LTS.} We also use Intel PAC D5005 Acceleration Card with an Intel Stratix 10 GX FPGA~\cite{d5005-url} to reproduce the results of HEXL-FPGA. For all accelerators that require the GPU as their target hardware, we use NVIDIA V100 GPU with $32$GB RAM for evaluation~\cite{v100}. \bluehighlight{All the experiments are evaluated in a docker environment with the docker version 20.10.21.}


\subsection{Evaluation Results}
\label{sec:eval-results}
\begin{figure}[t]
\centering
\includegraphics[width=0.6\columnwidth]{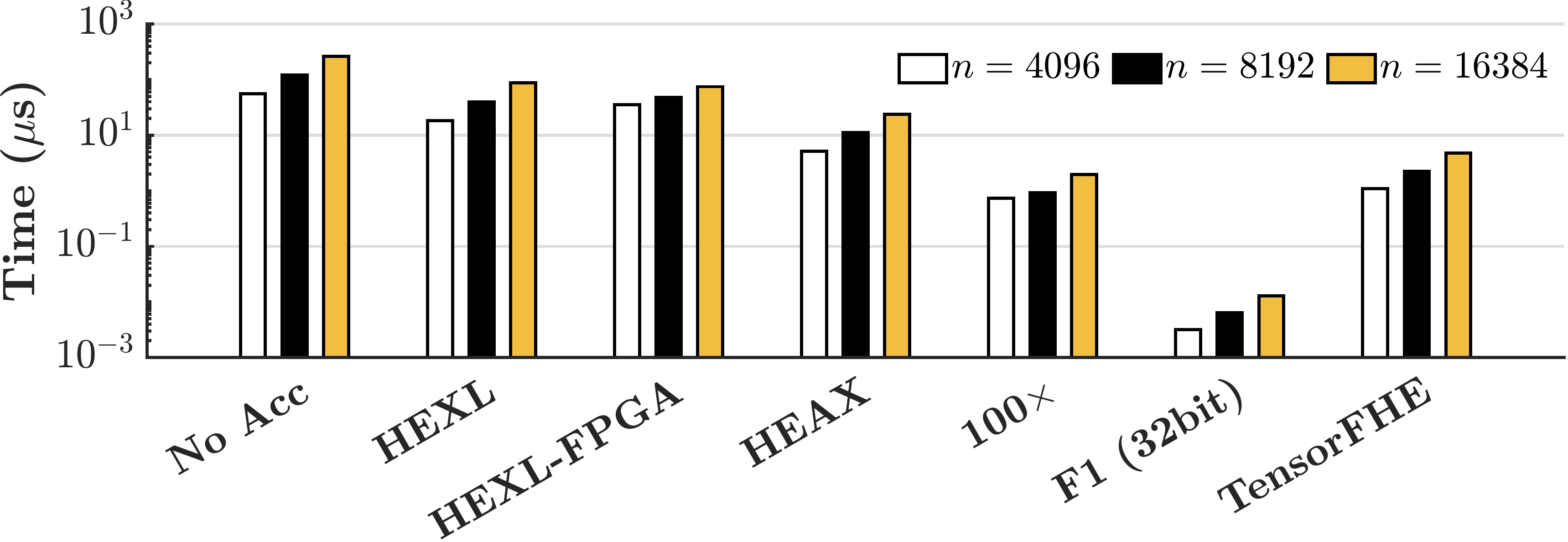}
\caption{Performance of NTT. The performance of HEXL, HEXL-FPGA and \ckksgpu is measured on our testbed while the performance results of HEAX and F1 are from their original paper.}
\label{fig:eval-ntt}
\end{figure}

\parab{NTT/FFT.} First, we will evaluate the performance of word-wise FHE accelerators. Figure~\ref{fig:eval-ntt} shows the NTT performance of HEXL, HEXL-FPGA, \ckksgpu, HEAX, F1 and TensorFHE. Please note that the performance of HEXL, HEXL-FPGA and \ckksgpu is measured on our testbed, while the performance results of HEAX, F1 and TensorFHE are from their original paper. We also run SEAL~\cite{seal-url} without any accelerators to demonstrate the baseline performance~(denoted as No Acc in the Figure). In this evaluation, we use three settings of $n$, \ie, $n=4096,8192,16384$. Similar to HEAX, we set the bit-width of polynomial coefficients in NTT to $52$ for evaluation~\cite{heax}. But F1 chooses the bit-width of $32$ as it is the largest word size in F1. Theoretically, the performance of F1 should be slightly worse if the bit-width is $52$.

We mainly have the following observations. 1) when the FHE accelerators leverage more advanced hardware technologies, the performance is largely improved. For example, the CPU-based accelerator, HEXL, can only achieve up to $3.0\times$ acceleration ratio, while ASIC-based accelerators, F1, can achieve up to $20546.9\times$ acceleration ratio. 2) contradicting our common wisdom, specific-designed hardware-based solutions do not always yield better performance than general hardware-based ones. For example, HEXL-FPGA and HEAX cannot achieve a better acceleration ratio over \ckksgpu and TensorFHE. The core reason is that HEXL-FPGA and HEAX adopt FPGA as their hardware platform, which suffers from the aforementioned problems such as limited programmable resources and low working frequency. Precisely, V100 GPU in our evaluation has the peak performance of $\sim 250$ INT8 TOPS with tensor core~\cite{v100}, which is $\sim 10\times$ better than Stratix 10 FPGA~\cite{stratix10} used in HEAX and HEXL-FPGA.

\begin{figure}[t]
\centering
\includegraphics[width=0.6\columnwidth]{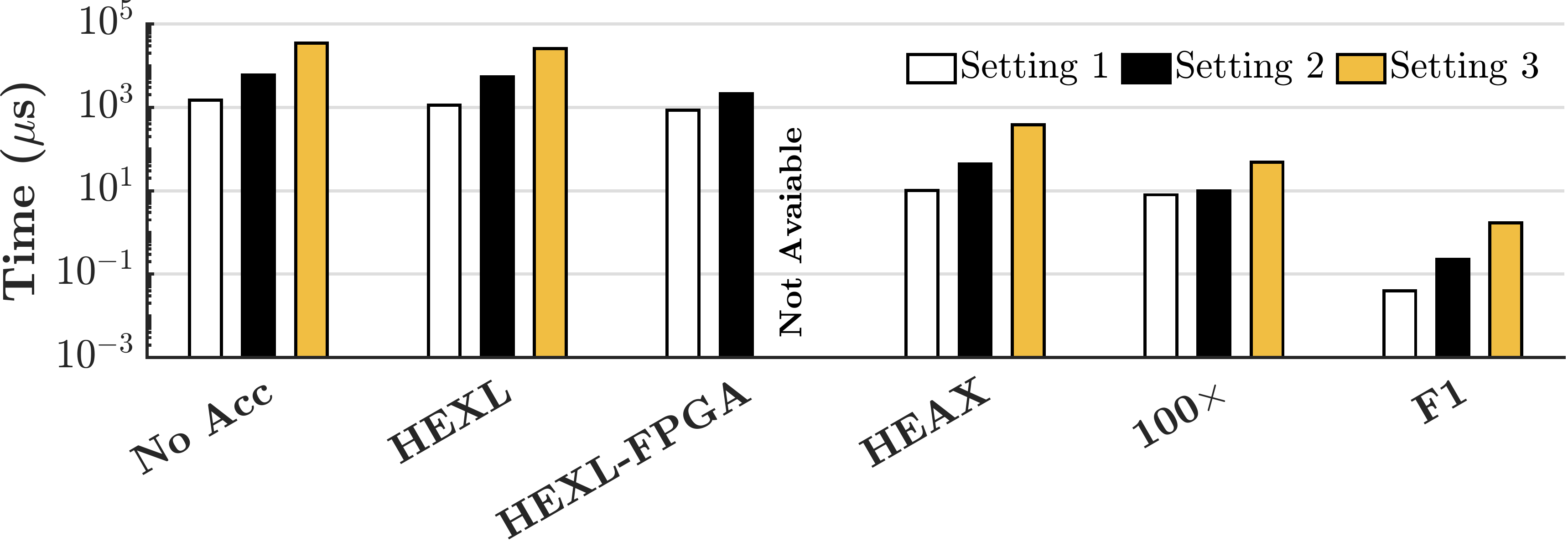}
\caption{Performance of key-switching. The performance of HEXL, HEXL-FPGA and \ckksgpu is measured on our testbed while the performance results of HEAX and F1 are from their original paper. HEXL-FPGA does not support Setting3.}
\label{fig:eval-key-switching}
\end{figure}

\begin{figure}[t]
\begin{minipage}[b]{0.48\columnwidth}
\centering
\includegraphics[width=0.6\textwidth]{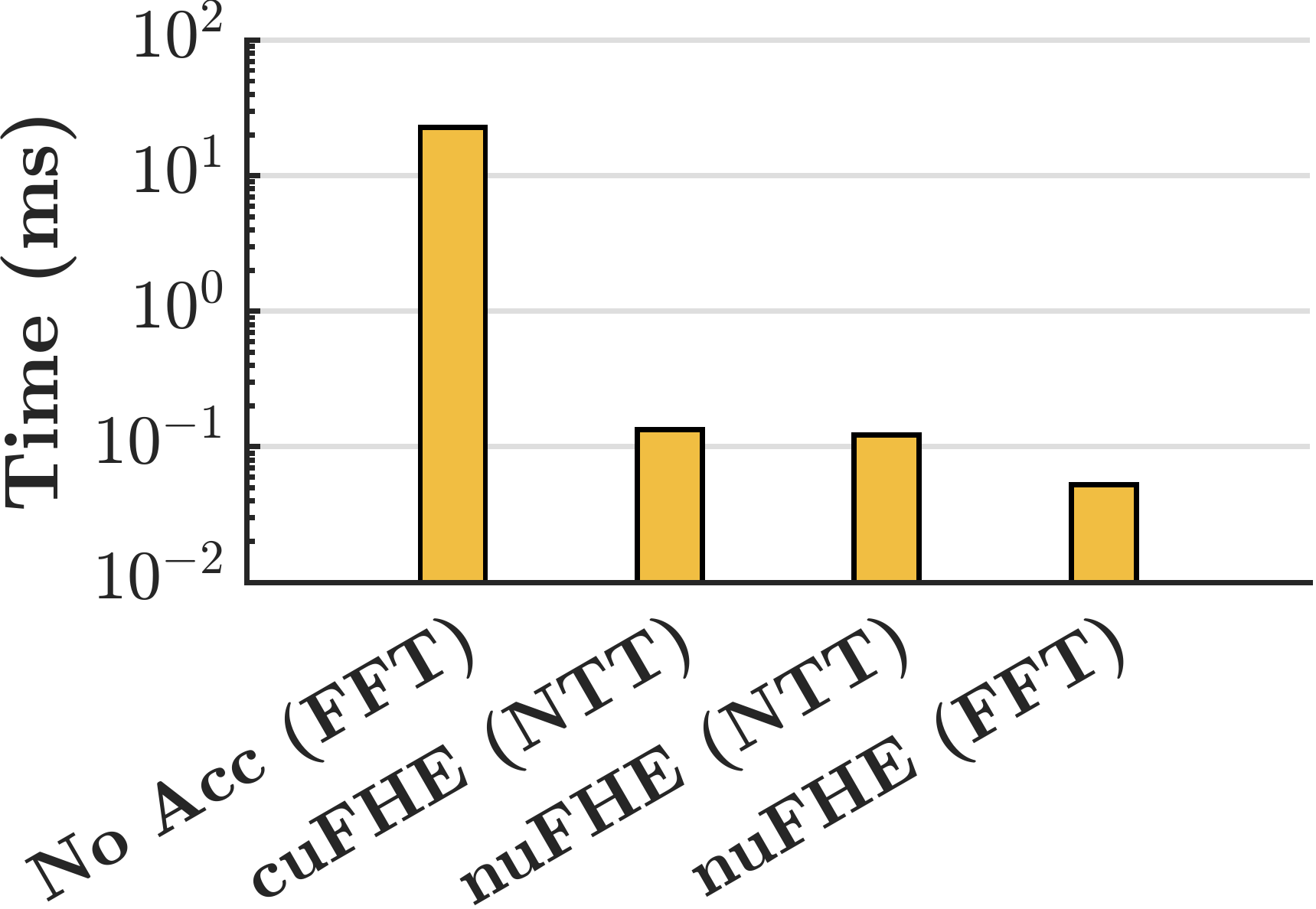}
\caption{Performance of NAND gate with NTT/FFT.}
\label{fig:eval-tfhe}
\end{minipage}
\hfill
\begin{minipage}[b]{0.48\columnwidth}
\centering
\includegraphics[width=0.6\textwidth]{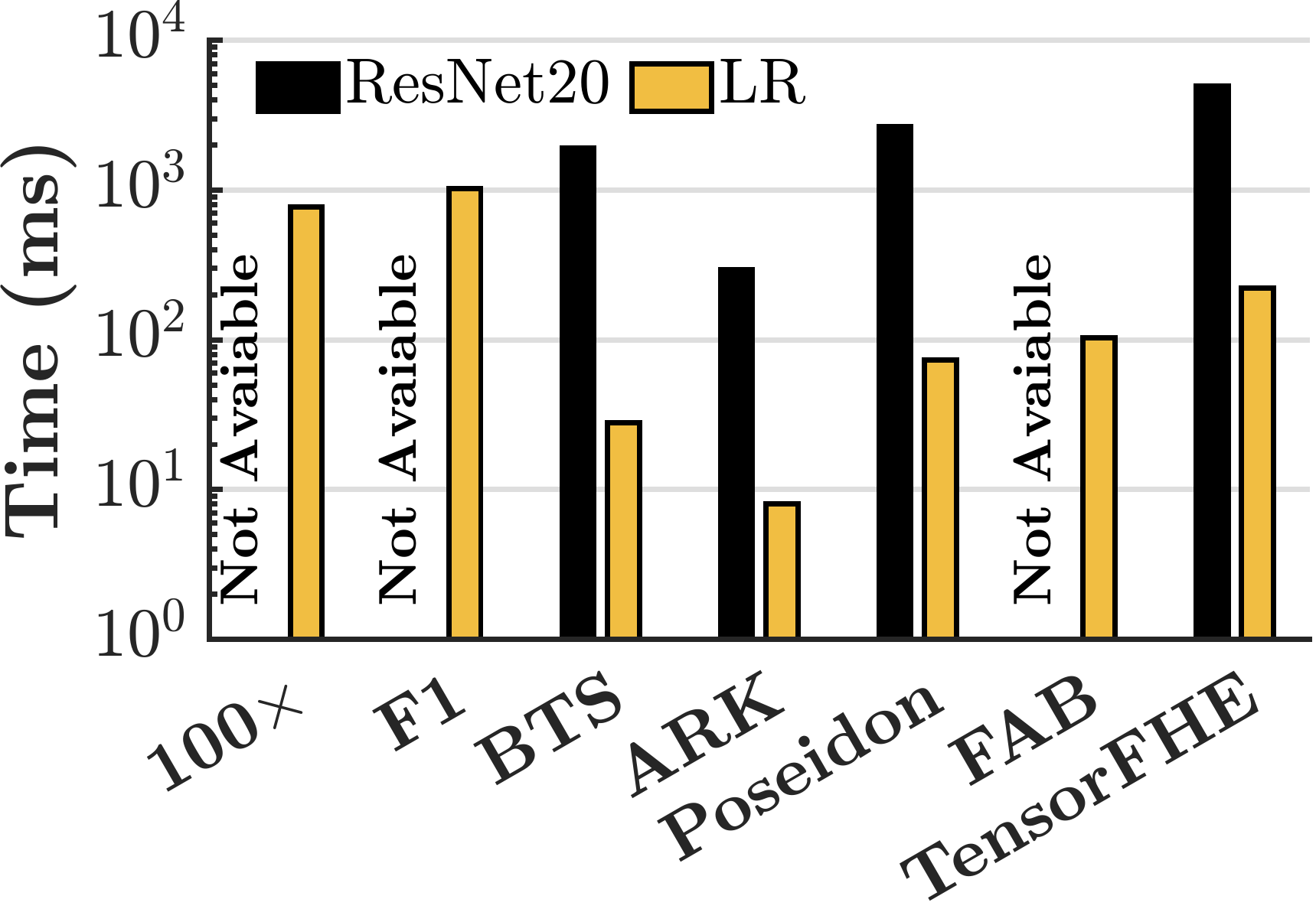}
\caption{Performance of end-to-end applications.}
\label{fig:eval-end-to-end}
\end{minipage}
\hfill
\end{figure}

Figure~\ref{fig:eval-tfhe} shows the performance of NAND gate achieved by cuFHE and nuFHE on our testbed. The NAND gate is a typical example that includes bootstrapping in bit-wise FHE schemes, such as TFHE. Polynomial multiplication in TFHE can be accelerated with either NTT or FFT, and we mark the schemes used by different libraries in the Figure. For the baseline~(No Acc), we run the TFHE software libraries~\cite{tfhe-url}.

From our evaluation, we can observe that nuFHE and cuFHE can achieve $170.6\times$ and $186.2\times$ acceleration with NTT implementation, respectively. Moreover, nuFHE also supports using FFT to accelerate the NAND gate, which can achieve up to $432.0\times$ acceleration.

\parab{Key-switching.} In our evaluation, we use three settings. \\
Setting1: $n=4096$, ~$L=1$, $\log(P\cdot Q)=109$, ${\rm dnum}=2$, \\
Setting2: $n=8192$, ~$L=3$, $\log(P\cdot Q)=218$, ${\rm dnum}=4$, \\ 
Setting3: $n=16384$, $L=7$, $\log(P\cdot Q)=438$, ${\rm dnum}=8$. \\
Figure~\ref{fig:eval-key-switching} shows the results. We have similar results as previous FFT/NTT experiments. Worth noting, HEXL-FPGA cannot perform key-switching with the most complicated setting, \ie setting3, which confirms the potential drawback as discussed in \S\ref{sec:hexl-fpga}.

\parab{End-to-end deep applications.} Figure~\ref{fig:eval-end-to-end} shows the end-to-end performance over two applications achieved by 6 latest accelerators, \ie F1, BTS, ARK, Poseidon, FAB and TensorFHE. Since these solutions are not open-sourced, we use the results from their original papers. 
We do not include the results of ResNet20 with F1, \ckksgpu and FAB since they are not presented in the original papers. The results reveal that: 1) As discussed in~\S\ref{sec:f1}, in a task requiring large multiplicative depth, F1 shows critical performance deficiencies despite its significant performance of NTT. For example, it takes F1 $1024$ms to execute a single iteration of LR, which is even slower than \ckksgpu~($775$ms), Poseidon~($73$ms), FAB~($103$ms) and TensorFHE~($222$ms);  2) For deep benchmarks, ASIC designs that support efficient bootstrapping achieves a significant increase in performance compared to FPGA and GPU designs. For example, BTS is $2.6\times$ and $7.8\times$ faster than FAB and TensorFHE with LR, respectively. As ARK proposes further improvement over bootstrapping, including algorithm optimization and architecture co-design, it achieves better performance when bootstrapping dominates the workflow. In ResNet20, where bootstrapping takes up to $76.2\%$ of the total workload, ARK achieves a $6.5\times$ speedup compared to BTS.

\bluehighlight{\parab{End-to-end shallow applications.} Recent works like  BTS, ARK, Poseidon, \etc, focus on the acceleration of deep workloads and do not cover much discussion of shallow applications. Since shallow applications are also widely used in real world, here, we briefly discuss the performance of F1 in the shallow workloads. In shallow applications, F1 achieves sufficient acceleration as bootstrapping is not required. In LoLa-CIFAR with unencrypted model weights, F1 achieves $5,011\times$ speedup compared to the CPU. In LoLa-MNIST, F1 is $17,412\times$ and $15,086\times$ faster than CPU when the model weights is unencrypted and encrypted, respectively. Subsequent works like CraterLake don't have a significant advantage over F1 in shallow tasks as they allocate a significant amount of hardware resources to operations specific to deep applications, such as bootstrapping.} 

\bluehighlight{
\parab{Impact of On-chip storage.} The performance of ASIC-based FHE accelerators can be significantly influenced by the size of on-chip memory storage~\cite{craterlake, bts, f1, ark}. Optimal acceleration is achieved by storing data in on-chip memory to minimize frequent data input/output between on-chip and off-chip memory. However, FHE schemes, particularly during key-switching operations, require a large amount of data, such as ciphertexts and switching keys. If the on-chip memory cannot accommodate all of this data, extensive data communication and exchange occur, resulting in a significant portion of the execution time.

\begin{figure}[t]
\centering
\includegraphics[width=0.6\columnwidth]{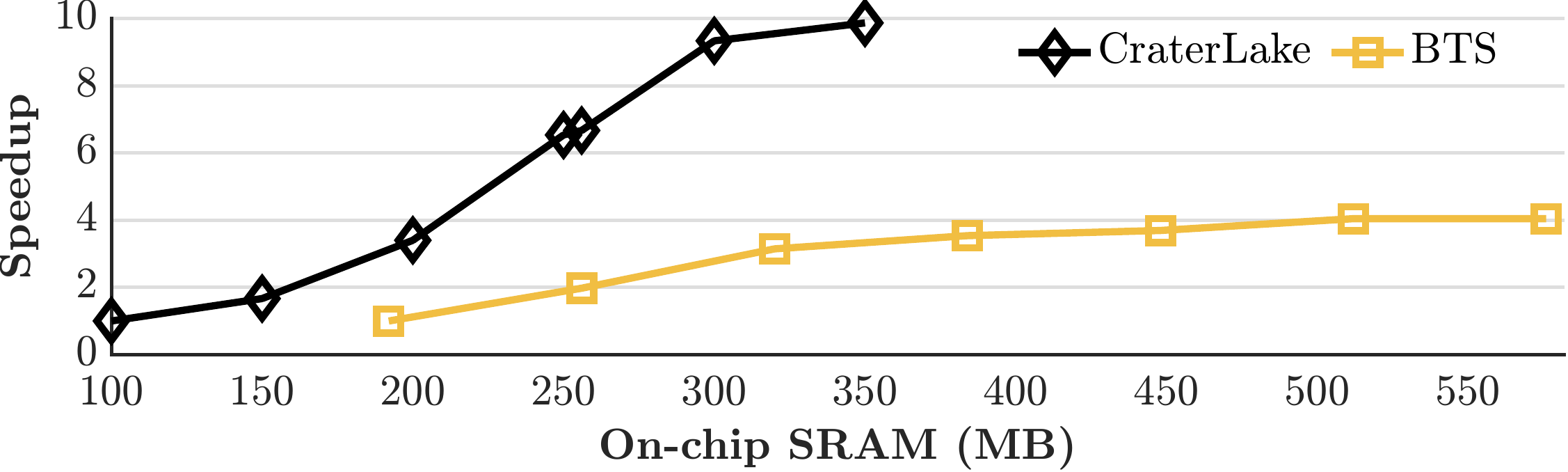}
\caption{Performance of Bootstrapping over increasing on-chip SRAM volume.}
\label{fig:eval-bootstrapping-sram}
\end{figure}

Figure~\ref{fig:eval-bootstrapping-sram} illustrates the impact of increasing on-chip SRAM memory on bootstrapping performance using CraterLake~\cite{craterlake} and BTS~\cite{bts}. These accelerators exhibit different internal designs and implementations, leading to varying behaviors as the volume of SRAM increases. However, a general trend can be observed: performance initially improves and then stabilizes with additional on-chip SRAM. This is because larger on-chip SRAM can store more intermediate data, reducing the frequency of costly data communication, even with high-speed HBM. Nevertheless, once the on-chip SRAM reaches a sufficient size where no further data exchange is required, additional SRAM does not contribute to further performance gains.
}

\section{Discussion on Future Directions}
\label{sec:discussion}
Inspired by the above qualitative ($\S$\ref{sec:fhe_acc_sok}) and quantitative ($\S$\ref{sec:evaluation}) studies, in this section we discuss some future directions of designing and implementing FHE accelerators. 

\subsection{Application-driven Design Approach}
Most existing FHE accelerators are trying to build increasingly powerful accelerators by leveraging more advanced architecture technologies, such as better semiconductor manufacturing processes~\cite{heax,f1,craterlake,bts,ark}. While such design methods can largely improve the performance of FHE schemes, they also become more unaffordable, especially for these ASIC-based solutions.

We envision that a future opportunity for FHE accelerators could be the application-driven design approach. The reason is that for some practical FHE applications, the performance is determined not only by the accelerators with extremely high FHE computing performance, but also by the whole architecture stack that customizes with the application. Table~\ref{tab:application-specific-acc} provides a brief overview of these works.
For example, FHE is widely adopted in private set intersection~(PSI), and a recent work called INSPIRE co-designs the storage architecture with the FHE accelerator to improve the end-to-end performance of the PSI application by minimizing the overhead of data movement between the storage controller and FHE accelerators~\cite{inspire}.
Cheetah~\cite{cheetah} and FxHENN~\cite{fxhenn} are two works targeting accelerating HE-based CNN inference applications by co-designing CNN inference workflows and HE algorithms.
We believe such an application-driven design approach could better balance the application requirements, costs, and design difficulties, providing a promising direction for the future.

\subsection{Supporting both Word-wise \& Bit-wise FHE Schemes}
As discussed in \S\ref{sec:fhe_background}, word-wise FHE is more suitable for polynomial evaluation, while bit-wise FHE is preferred for non-polynomial evaluation. However, real-world applications, such as machine learning tasks, require both polynomial and non-polynomial evaluations to be effective. Moreover, ciphertexts switch between the two forms~(\eg, CKKS and FHEW in \cite{pegasus}) is extremely complicated, thus time-consuming. Therefore, designing a FHE accelerator that can simultaneously support both word-wise and bit-wise FHE schemes, and further perform efficient ciphertext switch and evaluation is important for real-world applications, which is worthy of future investigation.

\subsection{Enhanced Software/Hardware Co-design}
\label{soft-hardware-co-design}
In recent works, the software has been considered an important part of the design. Some works, such as F1~\cite{f1}, CraterLake~\cite{craterlake}, use compilers for a software/hardware co-design solution. However, they do not provide a comprehensive description of the software design. We believe fully functional compilers should be co-designed with general FHE compilers, such as EVA~\cite{eva}. As the FHE compiler SoK paper~\cite{sok_fhe_compiler} indicates, general FHE compilers can optimize FHE programs based on the cost model of FHE schemes. We believe these general FHE compilers could be integrated with FHE accelerator compilers for improved end-to-end performance by considering a cost model from a hardware-level perspective, which points to a potential future direction.

\subsection{From Scale-up to Scale-out}
Current FHE acceleration solutions mostly focus on scale-up, \ie, improving the performance of a single accelerator vertically. However, scale-out, \ie, connecting multiple FHE accelerators via networking horizontally, should also be an effective way to further improve the performance of FHE applications. As discussed in \S\ref{sec:fhe_inefficiency}, the data inflation problem has posed a challenge to efficient data movement between on-chip and off-chip memories. This challenge also exists when we connect multiple FHE accelerators via networking in a scale-out manner. Thus, FHE accelerator and network co-design, \eg, integrating both FHE acceleration functions and high-performant networking controllers on the same chip, should be a potential research direction in the future.

\begin{table}[t]
\footnotesize
\centering
\begin{tabularx}{0.6\columnwidth}{l Y Y Y}
\toprule
\multirow{2}{*}{\bf Name} &
\multirow{2}{*}{\shortstack[c]{\bf Employed \\ \bf Hardware}} &
\multirow{2}{*}{\shortstack[c]{\bf Supported \\ \bf Application}} &
\multirow{2}{*}{\bf HE Schemes} \\
& & & \\
\midrule
INSPIRE~\cite{inspire} & ASIC & PIR & BFV \\
Cheetah~\cite{cheetah} & ASIC & CNN Inference & BFV \\ 
FxHENN~\cite{fxhenn} & FPGA & CNN Inference & CKKS \\
\bottomrule
\end{tabularx}
\caption{Application-specific FHE Accelerators.}
\label{tab:application-specific-acc}
\end{table}

\subsection{Accelerating NTRU-based Schemes}
\label{sec:ntru-acc}
In this paper, we do not cover accelerators for traditional NTRU-based solutions in details since \bluehighlight{(1) there are only few literatures targeting at designing accelerators for NTRU-based FHE schemes and} (2) these NTRU-based schemes were believed to be vulnerable to attacks and thus impractical~\cite{ntru-attack}. However, recently various modern NTRU-based solutions have been proposed to overcome its original security problems~\cite{final}. Moreover, NTRU-based solutions have the advantages of low memory consumption and fast computation. Therefore, we believe that designing practical accelerators for modern NTRU-based solutions also deserves future exploration.

\bluehighlight{
\begin{table}[t]
\small
\begin{tabularx}{0.6\linewidth}{l Y Y}
\toprule
\bf Employed Hardware & \bf AES~(s) & \bf Prince~(s) \\
\midrule
CPU~\cite{ntru-prince} & 55~(baseline) & 3.3~(baseline) \\
GPU~\cite{ntru-gpu} & 7.3~(7.5$\times\uparrow$) & 1.28~(2.58$\times\uparrow$)  \\
ASIC~\cite{ntru-asic} & 0.44~(125$\times\uparrow$) & 0.05~(66$\times\uparrow$)\\
\bottomrule
\end{tabularx}
\caption{\bluehighlight{Accelerators for NTRU-based FHE schemes.}}
\label{tab:ntru-performance}
\end{table}

We provide a brief summary of the performance of existing accelerators for NTRU-based schemes using GPUs \cite{ntru-gpu} and ASICs \cite{ntru-asic}. Following the conventions of these works, we use the homomorphic evaluation of AES and Prince as the performance metric \cite{ntru-prince}. The CPU implementation in \cite{ntru-prince} serves as the baseline for comparison. Since these NTRU-based accelerators are not open-sourced, we directly report the data from \cite{ntru-prince, ntru-gpu, ntru-asic}, and the results are presented in Table~\ref{tab:ntru-performance}. Compared to the CPU implementation, the GPU and ASIC-based accelerators show performance improvements ranging from $2.6\times$ to $7.3\times$ and $66\times$ to $125\times$, respectively.

One observation is that the acceleration ratio of NTRU-based accelerators, when compared to accelerators designed for CKKS/BFV/TFHE in \S\ref{sec:sok-survey}, is relatively lower. For instance, CraterLake can outperform the CPU by five orders of magnitude, while the latest accelerator for NTRU-based schemes \cite{ntru-asic} achieves only three orders of magnitude speedup. We identify two potential reasons for this discrepancy: (1) NTRU-based schemes themselves are reported to be faster than RLWE-based FHE schemes, leaving less room for further acceleration; (2) The latest ASIC-based accelerator for NTRU schemes is currently simpler than the accelerators discussed in \S\ref{sec:sok-survey}, indicating potential for further improvements.
}
\section{Conclusion}
\label{sec:conclusion}
\bluehighlight{
This paper presents a comprehensive systematization of knowledge through qualitative and quantitative analysis of 14 existing fully homomorphic encryption (FHE) accelerators and their evolution process. We have identified four key trends in the development of these accelerators. First, there is a growing emphasis on leveraging advanced hardware to achieve better acceleration performance. Second, a software/hardware co-design approach is commonly used. Third, there is a focus on enhanced programmability. Fourth, the support for an unlimited number of multiplications has improved through better bootstrapping operation support. Our survey aims to provide researchers with a comprehensive understanding of the current state of FHE accelerators. Additionally, we discuss potential future directions for designing and implementing FHE accelerators, such as developing accelerators for NTRU-based FHE schemes and considering a scale-out methodology instead of scale-up. We hope these discussions will inspire the research community and illuminate the future development of FHE accelerators.
}

\newpage

\bibliographystyle{ACM-Reference-Format}
\bibliography{refs}

\appendix

\section{NTT/FFT}
\label{app:ntt}
Given a polynomial $A$ of degree $n$ as follows:

\begin{equation}
    A(x) = \sum_{j=0}^{n-1} a_j x^j
\label{eq:polynomial}
\end{equation}

The polynomial can be represented in the form of the vector of its coefficients as $\boldsymbol{a} = (a_1, a_2, ..., a_{n-1})$. While the addition operation of two polynomials $\boldsymbol{a}$ and $\boldsymbol{b}$~(\ie, element-wise addition of their coefficients vectors) is trivial, the multiplication of $\boldsymbol{a}$ and $\boldsymbol{b}$~(denoted as $\boldsymbol{a} \bigotimes \boldsymbol{b}$ in our paper) is time-consuming with the computation complexity of $O(n^2)$. Since polynomial multiplication is the fundamental operation in FHE schemes, \eg, encryption, homomorphic operations, \etc, such high computation complexity causes FHE schemes to be extremely inefficient.

To optimize polynomial multiplication, another representation of the polynomial --- point-value representation --- can be exploited. A polynomial of degree $n$ can be represented by $n$ distinct point-value tuples: $\{(x_0,y_0), (x_1,y_1), ... (x_{n-1}, y_{n-1})\}$, where $y_k=A(x_k)$ and $k\in[0,n)$. To achieve calculations between polynomials, \eg, $A$ and $B$, the same group of $x_k$ is chosen from both polynomials $A$ and $B$. Under point-value representation, both addition and multiplication between two polynomials of degree $n$ only require $n$ point-wise operations on $y_k$. Therefore, the computation complexity is reduced to $O(n)$.

However, the conversion from coefficient representation to point-value representation is still time-consuming and requires the calculation of $y_k=A(x_k)$ for every $x_k$. The time complexity of evaluating each $A(x_k)$ is $O(n)$, thus leading to overall time complexity of $O(n^2)$ for $k\in [0,n)$. The same complexity is needed for the inverse conversion.

To lower the time complexity of representation conversion, we choose special $x_k$ values. For a polynomial of degree $n-1$, we use the roots of an equation $\omega^n=1$ as $x_k$ values. The equation has $n$ roots, denoted as $e^{2\pi ik/n}$, $k=0,1,...,n-1$. They are usually written in simplified form as $\omega_n^k=e^{2\pi ik/n}$, which is the power of $\omega_n=e^{2\pi i/n}$. With these roots, we can write Equation~\ref{eq:polynomial} as 

\begin{equation}
    A_k = A(\omega_n^k) = \sum_{j=0}^{n-1} a_j \omega_n^{kj}, k \in [0,n)
\label{eq:dft}
\end{equation}

Equation~\ref{eq:dft} is called Discrete Fourier Transform (DFT). Our goal is to accelerate the transformation with the property of $\omega_n$.

For $\omega_n$, we have the following theorems:

\begin{equation}
    \omega_{2n}^{2k} = \omega_n^k
\label{eq:utility_of_root_theorem_1}
\end{equation}

\begin{equation}
    \omega_{n}^{k+\frac{n}{2}} = - \omega_n^k
\label{eq:utility_of_root_theorem_2}
\end{equation}


Given a polynomial $A(x)=a_0 + a_1 \cdot x^1 + a_2 \cdot x^2 + ... +a_{n-1} \cdot x^{n-1}$, we can re-arrange it to $A(x)=(a_0 + a_2 \cdot x^2 + a_4 \cdot x^4 + ... + a_{n-2} \cdot x^{n-2}) + x(a_1 + a_3 \cdot x^2 + a_5 \cdot x^4 + ... + a_{n-1} \cdot x^{n-2})$. Let $A^{[0]}(x) = a_0 + a_2 \cdot x^1 + a_4 \cdot x^2 + ... + a_{n-2} \cdot x^{\frac{n-2}{2}}$ and $A^{[1]}(x) = a_1 + a_3 \cdot x^1 + a_5 \cdot x^2 + ... + a_{n-1} \cdot x^{\frac{n-2}{2}}$. Then we have $A(x) = A^{[0]}(x^2) + xA^{[1]}(x^2)$.

Therefore, we can use $x_k=\omega_n^k$ to sample values. For $k=0,1,\cdots,n/2-1$, we have

\begin{equation}
A(\omega_n^k) = A^{[0]}(\omega_n^{2k}) + \omega_n^kA^{[1]}(\omega_n^{2k})
\end{equation}

Based on Theorems~\ref{eq:utility_of_root_theorem_1}, the equation transforms into 
\begin{equation}
A(\omega_n^k) = A^{[0]}(\omega_{\frac{n}{2}}^k) + \omega_n^kA^{[1]}(\omega_{\frac{n}{2}}^k)
\label{eq:CT-butterfly-upperhalf}
\end{equation}

Similarly, for $k+n/2$, we have:
\begin{equation}
A(\omega_n^{k+\frac{n}{2}}) = A^{[0]}(\omega_n^{2k+n}) + \omega_n^{k+\frac{n}{2}}A^{[1]}(\omega_n^{2k+n})
\end{equation}

Based on Theorems~\ref{eq:utility_of_root_theorem_1} and \ref{eq:utility_of_root_theorem_2}, the equation now becomes:
\begin{equation}
A(\omega_n^{k+\frac{n}{2}}) = A^{[0]}(\omega_{\frac{n}{2}}^k) - \omega_n^kA^{[1]}(\omega_{\frac{n}{2}}^k)
\label{eq:CT-butterfly-lowerhalf}
\end{equation}

Note that $k$ and $k+n/2$ have covered all integers ranging from 0 to $n-1$. Then the problem becomes a divide-and-conquer problem. The origin problem $A$ is split into two subproblems of $A^{[0]}$ and $A^{[1]}$ whose degrees of polynomial are $n/2$. It requires $n/2$ extra multiplications to recover $A$ from $A^{[0]}$ and $A^{[1]}$. After that, $A^{[0]}$ and $A^{[1]}$ can be further divided into smaller subproblems following the same scheme. The recursion ends when the degree of the polynomial in the subproblem is $1$. According to the master theorem in~\cite{intro_to_alg}, the time complexity can be reduced to $O(n \log n)$.

The aforementioned process is called Fast Fourier Transform~(FFT). When the degree of polynomials is $n$, we refer to the transform as an $n$-point FFT. The inverse operation of FFT is called iFFT, which can be implemented with a similar approach.

Number Theoretic Transforms~(NTT) is similar to DFT but works with the finite field. Different from DFT in Equation~\ref{eq:dft}, NTT is formulated as

\begin{equation}
    A_k = \sum_{j=0}^{n-1} a_j g_n^{kj} \mod p, k \in [0,n),
\label{eq:ntt}
\end{equation}
where $p$ is a prime, and $g_n$ is the primitive n-th root of unity modulo $p$, which satisfies the following theorems:

\begin{equation}
    g_{2n}^{2k} \equiv g_n^k \mod p
\label{eq:utility_of_root_theorem_3}
\end{equation}

\begin{equation}
    g_{n}^{k+\frac{n}{2}} \equiv - g_n^k \mod p
\label{eq:utility_of_root_theorem_4}
\end{equation}

With \bluehighlight{Theorems}~\ref{eq:utility_of_root_theorem_3} and~\ref{eq:utility_of_root_theorem_4}, NTT can be accelerated following the same scheme we discussed in FFT, except for the multiplication and addition/subtraction, which need to be replaced by modular multiplication and modular addition/subtraction, respectively.

\section{4-step FFT/NTT Algorithm}
\label{app:4-step}
\subsection{Introduction to 4-step FFT/NTT Algorithm}
\label{app:intro-4-step}
In this section, we use the example of an $n$-point FFT to illustrate the workflow of the 4-step FFT/NTT algorithm. Recall the original FFT~(DFT) in Equation~\ref{eq:dft}. In FHE schemes, $n$ is a power of two and can be expressed as the product of two numbers $n=R\cdot C$. Based on this property, we consider the $n$ input numbers $\{a_j, j\in [0,n)\}$ as an $R\times C$ matrix. In this way, the workflow of the 4-step FFT is as follows.

\begin{icompact}

\item[1.] Transpose the $R\times C$ input matrix and get a new $C\times R$ matrix. Perform FFT on each row of the $C\times R$ matrix~(\ie, $C$ independent $R$-point FFTs). We let matrix $\boldsymbol{A'}$ denote the results of FFTs.

\item[2.] Transpose matrix $\boldsymbol{A'}$ and get a new $R\times C$ matrix $\boldsymbol{A''}$.

\item[3.] Generate a $R\times C$ twisting factor matrix $\boldsymbol{F}=[F_{i,j}]$, where $F_{i,j}=\omega_n^{ij}$. Then perform dyadic multiplication between matrix $\boldsymbol{A''}$ and matrix $\boldsymbol{F}$. Let matrix $\boldsymbol{A'''}$ denote the multiplication result.

\item[4.] Perform FFT on each row of the $R\times C$ matrix $\boldsymbol{A'''}$ (\ie, $R$ independent $C$-point FFTs). Transpose the result of FFT and get a new $C\times R$ matrix, denoted as $\boldsymbol{A}$.

\end{icompact}

$\boldsymbol{A}$ is the final result of the $n$-point FFT with some differences in the placement order of the result numbers from the original FFT algorithm. The 4-step NTT algorithm follows a similar workflow with modular operations between integers. Since the order of coefficients does not affect the correctness of element-wise polynomial addition and multiplication, the 4-step FFT/NTT algorithm can be easily applied in FHE schemes to replace the original FFT/NTT algorithm. Actually, we could skip the transpose in the first step by storing the input numbers in column-major order, and skip the transpose in the fourth step because of the order-independent element-wise operations.






\subsection{Comparison between the 4-step FFT/NTT Algorithm and the Original FFT/NTT Algorithm}
\label{app:comp-4-step}
This section will show the pros and cons of the 4-step FFT/NTT algorithm by comparing it with the original FFT/NTT algorithm.

\subsubsection{Efficient and Practical Parallelism}
As mentioned in \S\ref{sec:ntt_complexity}, the original FFT/NTT algorithm can hardly be accelerated via parallelism because of strict data dependency and high computation resource consumption. The 4-step FFT/NTT algorithm alleviates both problems.

\parab{Less Data Dependency:}  According to the aforementioned workflow, the 4-step FFT/NTT algorithm decomposes the $n$-point FFT/NTT operation into $C$ independent $R$-point FFT/NTTs in the first step, a global transpose in the second step, $n$ independent multiplications in the third step, and $R$ independent $C$-point FFT/NTTs in the last step. Although the four steps should be executed one by one, the parallelism in these steps~(except for the second step) is fully exploited as these independent operations can be performed simultaneously. Since $R$ and $C$ can be quite large (\eg, when $n=65536$, $R=C=256$), the parallelism is high enough to fully utilize the performance capacity of hardware accelerators. 
Actually, the global data dependency problem in the original FFT/NTT algorithm still exists in the second step when transposing the matrix.
Consequently, a large transpose network is required. Recent works prefer such a design since 1) the transpose network is much easier to design than FFT/NTT due to its fixed workflow; 2) the transpose network can be reused in other operations such as permutation~\cite{f1}, which improves the resource utilization. 

\parab{Low Resource Consumption:} Compared to the original FFT/NTT, the total computation workload in the 4-step FFT/NTT is not reduced. But the original algorithm is decomposed into multiple smaller ones with much fewer input numbers~(we choose $R$ and $C$ close to $\sqrt{n}$ in many cases). Therefore, instead of designing a fully pipelined $n$-point FFT/NTT circuit, we can combine multiple small pipelined FFT/NTT units~(\ie, $R$-point and $C$-point FFT/NTT units) to achieve the same arithmetic function, reducing the hardware design complexity.
To reduce hardware resource consumption, we can reuse 
$R$-point FFT/NTT units rather than implement all the units as dedicated ones~(\ie, $R$ $C$-point FFT/NTT units and $C$ $R$-point FFT/NTT units). Therefore, the 4-step FFT/NTT algorithm requires much fewer hardware resources to achieve pipeline parallelism.

\subsubsection{Increased Memory Overhead}
Although promising, the 4-step FFT/NTT also causes increased memory overhead due to pre-computed parameters. In addition to the input data and calculation results, we need to store the twiddle factors for $n$-point FFT/NTT in the original algorithm. Larger $n$ leads to more twiddle factors. Although the number of twiddle factors is reduced in the 4-step FFT/NTT algorithm since we only need twiddle factors for $R$-point and $C$-point FFT/NTTs, we need to store the twisting factors introduced in the third step of the 4-step FFT/NTT algorithm. The size of twisting factors in the 4-step FFT/NTT algorithm is close to the size of twiddle factors in the original algorithm, making the requirements of total storage even more strict. To address the problem, recent works generate twiddle and twisting factors on the fly rather than cache all the factors in the memory~\cite{bts,craterlake}.

\end{document}